\documentclass[conference]{IEEEtran}
\IEEEoverridecommandlockouts
\usepackage{amsmath}
\usepackage{algorithmic}
\usepackage{graphicx}
\usepackage{textcomp}
\usepackage{xcolor}
\usepackage{hyperref}
\usepackage{enumitem}
\usepackage{subcaption}
\usepackage{listings}
\usepackage{multirow}
\usepackage{pifont}
\usepackage{xspace}
\usepackage{float}
\floatstyle{plaintop}
\restylefloat{table}
\usepackage{subcaption}
\usepackage{comment}
\usepackage{booktabs}
\usepackage{cuted}

\usepackage{makecell}

\usepackage[dvipsnames]{xcolor}

\definecolor{webblue}{rgb}{0,0,.7}

\lstdefinestyle{SQLStyle}{
  language=SQL,
  basicstyle={\scriptsize\ttfamily},
  breaklines=fullflexible,
  frame=single, 
  frameround=tttt, 
  backgroundcolor=\color{gray!5}, 
  rulesep=1pt,
  numbers=none,
  keepspaces=true,
  showstringspaces=false,
  captionpos=b,
  aboveskip=10pt,
  belowskip=0pt,
  numberstyle=\tiny\color{gray},  
  keywordstyle=\color{webblue},
  commentstyle=\color{gray},
  keywords=[2]{LATERAL, UNNEST, APPLY},
  keywordstyle=[2]\color{webblue},
}

\newcommand{\mysimplenote}[1]{{#1}}

\newcommand{\revisionfigure}[1]

\definecolor{Salmon}{HTML}{FA8072}

\newcommand{\change}[1]{{\mysimplenote{#1}}}
\newcommand{\shorten}[1]{{\color{black} \mysimplenote{#1}}}

\newcommand{\mypar}[1]{\noindent\textbf{#1.}\xspace}

\newcommand{\sysname}{Exqutor\xspace}
\newcommand{\benchmarkname}{Vector-augmented SQL analytics\xspace}

\newcounter{point}[reviewer]

\def\BibTeX{{\rm B\kern-.05em{\sc i\kern-.025em b}\kern-.08em
    T\kern-.1667em\lower.7ex\hbox{E}\kern-.125emX}}
    
\begin{document}

\title{Exqutor: Extended Query Optimizer for Vector-augmented Analytical Queries}

\author{
    \IEEEauthorblockN{
        Hyunjoon Kim\IEEEauthorrefmark{1}, 
        Chaerim Lim\IEEEauthorrefmark{1}, 
        Hyeonjun An\IEEEauthorrefmark{1}, 
        Rathijit Sen\IEEEauthorrefmark{2}, 
        Kwanghyun Park\IEEEauthorrefmark{1}\textdaggerdbl
    }
    \IEEEauthorblockA{
        \IEEEauthorrefmark{1}Yonsei University BDAI Lab, \IEEEauthorrefmark{2}Microsoft Gray Systems Lab
        \\\{wns41559, chaerim.lim, hyeonjun.an\}@yonsei.ac.kr, rathijit.sen@microsoft.com, kwanghyun.park@yonsei.ac.kr
    }
}

\maketitle

\begingroup
\renewcommand\thefootnote{\textdaggerdbl}
\footnotetext{\textnormal{Corresponding author.}}
\endgroup
\begingroup
\renewcommand\thefootnote{}
\footnotetext{Accepted to ICDE 2026; to appear in the \textit{42nd IEEE International Conference on Data Engineering}.}
\endgroup

\begin{abstract}
Vector similarity search is becoming increasingly important for data science pipelines, particularly in Retrieval-Augmented Generation (RAG), where it enhances large language model inference by enabling efficient retrieval of relevant external knowledge. As RAG expands with table-augmented generation to incorporate structured data, workloads integrating table and vector search are becoming more prevalent. However, efficiently executing such queries remains challenging due to inaccurate cardinality estimation for vector search components, leading to suboptimal query plans.

In this paper, we propose \sysname, an \textbf{\underline{ex}}tended \textbf{\underline{qu}}ery optimizer for vec\textbf{\underline{tor}}-augmented analytical queries. \sysname is a pluggable cardinality estimation framework designed to address this issue, leveraging exact cardinality query optimization techniques to enhance estimation accuracy when vector indexes (e.g., HNSW, IVF) are available. In scenarios lacking these indexes, we employ a sampling-based approach with adaptive sampling size adjustment, dynamically tuning the sample size to balance estimation accuracy and sampling overhead. This allows \sysname to efficiently approximate vector search cardinalities while minimizing computational costs. We integrate our framework into pgvector, VBASE, and DuckDB, demonstrating performance improvements of up to four orders of magnitude on vector-augmented analytical queries.
\end{abstract}


\section{Introduction}
\label{sec:intro}
\begin{figure*}
\centering
\includegraphics[width=0.99\linewidth, trim={0pt 1pt 0pt 0pt},clip]{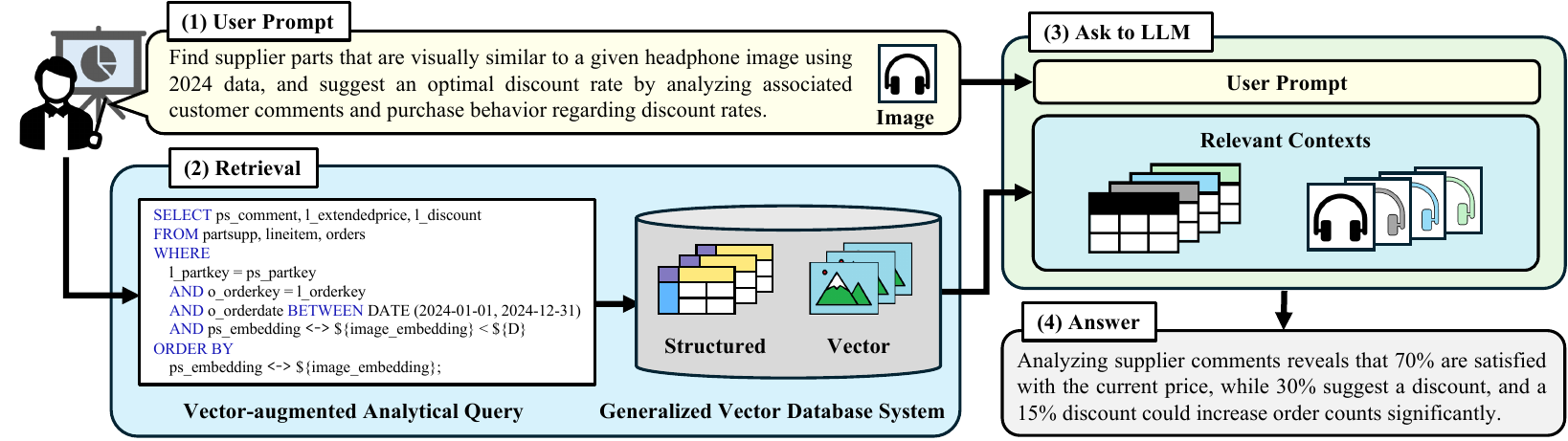}
\caption{Extended RAG pipeline integrating vector search with structured data. This four-stage pipeline retrieves structured and vector-based contexts to generate informed responses to user prompts. (1) The user provides a prompt, requesting optimal discount rate. (2) The user transforms the prompt into a VAQ to retrieve the relevant structured and vector data. (3) The retrieved contexts, along with the user prompt, are provided as input to the LLM. (4) The LLM generates a response, delivering analytical insights based on the combined structured and vector-based retrieval results.
}
\label{fig:pipeline}
\vspace{-6mm}
\end{figure*}

Retrieval-Augmented Generation (RAG) workflows have become essential for integrating Large Language Models (LLMs) into domains such as e-commerce, healthcare, and recommendation systems \cite{AnalyticDB, healthcare_rag_2, recommendation_rag_1}. By leveraging relevant external knowledge, RAG significantly enhances LLM performance and mitigates common challenges such as hallucinations and outdated information. In a traditional RAG pipeline, a vector similarity search (VSS) is a primary operator to retrieve the top-k most semantically similar documents, which are already embedded and saved in database systems. It can be performed either within specialized vector database systems optimized for vector data or within generalized vector database systems, which extend traditional relational databases with vector search capabilities.

\mypar{Advances in vector similarity search}
While traditional RAG relies on simple top-k retrieval over unstructured vector data, real-world analytical workloads increasingly demand the joint retrieval of both structured and unstructured data \cite{suv, AnalyticDB, ml_join, context-enhanced-realtional-operator}.
Consequently, significant advancements have been made in filtered vector search and vector range search. 
The filtered vector search augments vector similarity search with attribute-based filtering. This pattern captures many real-world applications, such as filtering by price and brand in e-commerce. The vector range search generalizes the simple top-k vector search by retrieving all vectors within a specified threshold from the target vector. It is commonly used to retrieve all sufficiently similar vectors, such as in plagiarism detection~\cite{singlestore}. Those two techniques can be used together to form more complex analytical queries like vector range search with filters.
Companies such as Alibaba \cite{AnalyticDB}, Apple \cite{apple}, Huawei \cite{andb}, and Microsoft \cite{vbase} have already restructured Approximate Nearest Neighbor (ANN) indexes and database systems to efficiently incorporate those types of queries.
Building on these developments, modern vector database systems are expected to support traditional query operators with VSS together \cite{Milvus, vbase, singlestore, qdrant, pgvector, clickhouse}.

\mypar{Vector-augmented analytical queries} 
Despite recent advances in attribute-based filtering and range-based vector search, existing work has assumed that vector embeddings and structured attributes are stored together within a single relation~\cite{singlestore, AnalyticDB, vbase, apple, Milvus}. However, in real-world deployments where the data scales in volume and schema complexity, this assumption becomes impractical \cite{blended_rag, scalable_springer, spider}. Modern database systems commonly partition data across multiple relations, requiring join operations to execute analytical queries on large datasets.

\begin{lstlisting}[style=SQLStyle, caption={An example VAQ generated for a user prompt in a RAG pipeline (see~\autoref{fig:pipeline}).}, label={lst:rag_query}]
  SELECT ps_comment, l_extendedprice, l_discount
  FROM partsupp, lineitem, orders
  WHERE
    l_partkey = ps_partkey
    AND o_orderkey = l_orderkey
    AND o_orderdate BETWEEN DATE (2024-01-01, 2024-12-31)
    AND ps_embedding <-> ${image_embedding} < ${D}
  ORDER BY
    ps_embedding <-> ${image_embedding};
\end{lstlisting}
 
To bridge this gap, we introduce Vector-augmented Analytical Queries (VAQs), which integrate vector similarity search with relational operators such as filters, aggregates, and joins. 
\autoref{lst:rag_query} presents an example VAQ that retrieves parts from many suppliers visually similar to a given product image, incorporating vector similarity search (\texttt{ps\_embedding <-> \$\{image\_embedding\}}) with relational joins, where \texttt{<->} denotes the Euclidean distance operator. The query applies a vector distance threshold $D$ to ensure that image embeddings close enough to the query embedding are selected.
\autoref{fig:pipeline} illustrates how data analysts leverage a RAG pipeline that integrates vector search with structured data to derive recommended discount rates based on supplier feedback and purchasing patterns \cite{nhq}. 
Generalized vector database systems, such as pgvector \cite{pgvector}, VBASE \cite{vbase}, and DuckDB \cite{duckdb}, are inherently better suited for executing VAQs, as they provide multi-relational queries such as complex filtering and aggregation. In contrast, specialized vector database systems~\cite{specialized-dbms} like Milvus \cite{Milvus} and Qdrant \cite{qdrant} are limited to single relation queries. This often necessitates coupling with an external RDBMS, which introduces additional system complexity and risks of data silos \cite{milvus_timeline}.

\mypar{Optimization opportunities} We observe that generalized vector database systems currently suffer from limited performance on VAQs. As shown in \autoref{sec:motivation}, this issue stems primarily from suboptimal query execution plans caused by inaccurate cardinality estimation, especially for VSS operators. Our investigation further reveals that several prominent generalized vector database systems rely on fixed default selectivity values when estimating VSS cardinality. This observation highlights a key optimization opportunity: improving VSS cardinality estimation can significantly enhance VAQ performance in generalized vector database systems.

\mypar{Contributions}
We propose \sysname, an open-source framework\footnote{\label{fn:github}\url{https://github.com/BDAI-Research/Exqutor}} designed to optimize query plans for VAQs by improving cardinality estimation of vector similarity searches. \sysname introduces Exact Cardinality Query Optimization (ECQO) for queries with vector indexes, executing Approximate Nearest Neighbor (ANN) searches during query planning to obtain accurate cardinality estimates. \change{Since ECQO can incur substantial computational overhead, \sysname addresses this challenge by visiting only tables with vector embeddings and reusing the obtained ANN results during execution.
For queries without vector indexes, \sysname employs an adaptive sampling-based approach that estimates predicate selectivity by evaluating similarity over a subset of data. This approach adaptively adjusts the sample size using momentum-based feedback \cite{momentum} and learning rate scheduling to balance estimation accuracy and planning efficiency.}
\change{By addressing the critical challenge of cardinality estimation for high-dimensional vector data, this work not only delivers practical performance gains but also opens up new avenues for advanced query optimization research for vector database systems.}

\change{
To the best of our knowledge, we introduce the \textit{first analytics benchmark that integrates both vector embeddings and structured data} to assess query optimization strategies for VAQs. 
In addition, \sysname is the \textit{first framework to explicitly address VAQs in real analytical scenarios}, highlighting and mitigating the effects of incorrect cardinality estimates on query performance.  
Our contributions are as follows:
}

\change{
\textbullet{} \textbf{Vector-augmented SQL analytics (\autoref{sec:benchmark})}:  
An analytical benchmark that extends TPC-H~\cite{tpc-h} and TPC-DS~\cite{tpc-ds_1} with widely used vector datasets by introducing embedding columns. 
This design enables systematic evaluation of VAQs under both classical relational and modern hybrid settings, and the benchmark is publicly available as open-source\textsuperscript{\ref{fn:github}}
}

\change{
\textbullet{} \textbf{Exact cardinality query optimization with vector index (\autoref{sec:ECQO})}:  
An optimization technique for VAQs with vector indexes, integrating ECQO into query planning by restricting ANN searches to vector tables. The results are reused at execution time without incurring redundant overhead.}

\textbullet{} \textbf{Adaptive sampling for cardinality estimation in VAQs without vector indexes (\autoref{sec:sampling})}:  
A sampling-based cardinality estimation approach for VAQs without vector indexes. \sysname employs adaptive sampling to improve accuracy by dynamically adjusting the sample size based on Q-error through a momentum-based adjustment mechanism and a learning rate scheduler.

\textbullet{} \textbf{Evaluation of vector-augmented SQL analytics with diverse datasets across multiple systems (\autoref{sec:eval})}:  
Comprehensive evaluation using vector-augmented SQL analytics with real-world vector datasets. \sysname achieves speedups of up to three orders of magnitude on pgvector, four orders of magnitude on VBASE, and 37.2{\small$\times$} on DuckDB. 
\section{Background}
\label{sec:background}

\mypar{Specialized vector database systems}
A specialized vector database system is a dedicated system designed for managing high-dimensional vector data, particularly the embeddings generated by machine learning models \cite{Vector-Database-Management}. Systems such as Milvus \cite{Milvus}, Pinecone \cite{pinecone}, Qdrant \cite{qdrant}, Manu \cite{Manu}, and ChromaDB \cite{chroma} are representative examples. These systems are architected with a strong focus on optimizing vector similarity search performance, incorporating advanced indexing techniques to accelerate approximate nearest neighbor search. While they deliver high efficiency in vector operations, their support for structured query capabilities remains limited \cite{survey-vector}. In practice, they mainly support filtered vector queries, which combine vector similarity search with attribute-based filters over metadata, and these queries are restricted to execution over a single collection.

\mypar{Generalized vector database systems}
Extending a traditional database system to support vector-type operators is referred to as a generalized vector database system. Examples include pgvector \cite{pgvector}, PASE \cite{PASE}, VBASE \cite{vbase}, and AnalyticDB \cite{AnalyticDB}, which are built on PostgreSQL as an extension, as well as DuckDB-VSS \cite{duckdb-vss}, which is based on DuckDB \cite{duckdb}. Other systems, such as SingleStore-V \cite{singlestore} and ClickHouse \cite{clickhouse}, are also built on relational database systems. Additionally, NoSQL-based systems including Vespa \cite{vespa}, Neo4j \cite{neo4j}, and Redis \cite{redis} can be classified in this category. These systems are not designed solely for vector operations and generally exhibit lower vector similarity search performance than specialized vector database systems \cite{survey-vector}. However, they provide robust support for managing a wide range of traditional data types.

\mypar{Nearest neighbor search and vector index}
Nearest Neighbor (NN) search is a core operation in vector similarity search, used to identify the most similar data points to a given query vector based on distance metrics such as cosine similarity or Euclidean distance. Two primary variants of this search exist: K-Nearest Neighbor (KNN) \cite{knn-1, knn-2, knn-3}, which offers exact results, and Approximate Nearest Neighbor (ANN), which improves efficiency by sacrificing some accuracy.
KNN performs an exhaustive comparison with all vectors in the dataset, guaranteeing precise results. However, its computational cost scales poorly with data size, making it impractical for large-scale or real-time applications.
In contrast, ANN reduces search cost by leveraging pre-built vector indexes that approximate the nearest neighbor computation. This trade-off between accuracy and performance makes ANN suitable for data-intensive workloads such as RAG and real-time recommendation systems. 
Most vector database systems support both KNN and ANN, allowing users to select the appropriate method depending on application needs. Efficient ANN search depends heavily on the underlying indexing technique. 
Common approaches include graph-based \cite{hnsw,graph,graph2,diskANN}, hash-based \cite{hash,hash3}, tree-based \cite{tree-aqp, tree-aqp-2, annoy}, and quantization-based methods \cite{pq, faiss, quantization}. Among them, HNSW \cite{hnsw} is one of the most widely adopted due to its high efficiency in high-dimensional spaces. HNSW builds a layered proximity graph and uses greedy traversal strategies to locate approximate neighbors, significantly reducing search latency while maintaining acceptable accuracy.

\mypar{Exact cardinality query optimization (ECQO)} 
Query optimizers often generate suboptimal plans, especially when cardinality estimation is challenging due to data skew and complex predicates \cite{ecqo_imma, ecqo_vec}. Vector similarity search exemplifies this challenge, as multi-dimensional spaces, large datasets, and query-dependent similarity thresholds complicate estimation \cite{kepler, ecqo_vec}. ECQO is the method that uses intermediate result sizes by actually executing queries during query plan generation rather than relying solely on simplifying assumptions and coarse metadata. Although this approach ensures accurate cardinality values for optimal plans, it has traditionally been considered an offline technique due to computational overhead \cite{ecqo_imma, ecqo_test}. This overhead arises from evaluating all relevant relations in a query. To address this, ECQO is selectively applied to critical expressions that are expected to significantly influence join ordering or plan cost estimation and used for relation pruning, reducing computational costs \cite{ecqo_relevent, ecqo_imma}.
This is particularly beneficial for multi-join queries, where cardinality estimation errors often degrade performance. In such scenarios, the additional overhead introduced by ECQO is justified by its considerable impact on the decrease in total execution time \cite{ce_important_microsoft, cardinality_important}.

\section{Motivation}
\label{sec:motivation}

VAQs can be executed on two types of database systems: specialized or generalized vector database systems. Specialized vector database systems, dedicated to managing vector data such as Milvus \cite{Milvus} and Qdrant \cite{qdrant}, support relational metadata storage and attribute-based filtering within a single collection. However, when handling a large number of structured attributes, these systems often require integration with an external RDBMS \cite{milvus_timeline}, resulting in increased complexity for development and query processing overhead \cite{AnalyticDB}. Consequently, they become less suitable for applications that require seamless and efficient management of a large volume of structured attributes alongside vector data.

\begin{figure}[t!]
    \centering
    \includegraphics[width=0.99\linewidth, trim={0pt 17pt 0pt 3pt},clip]{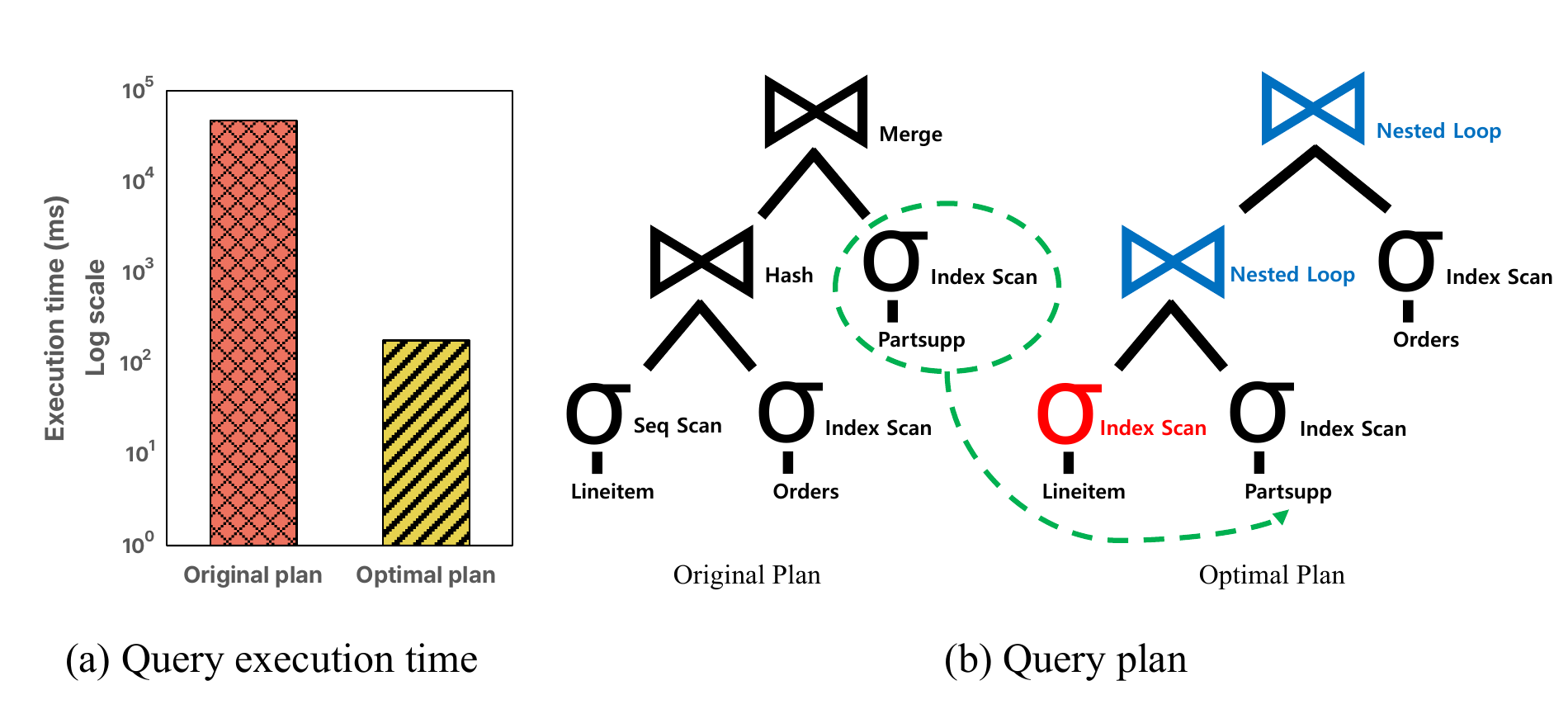}
    \caption{Execution time and generated query plan for the VAQ in \autoref{lst:rag_query} on pgvector. The optimal plan is generated by query optimizer with true cardinality of vector similarity search. The ps\_embedding column in the partsupp table (80M) has vector embeddings from the DEEP dataset.}
    \vspace{-2mm}
    \label{fig:motivation}
\end{figure}

\shorten{
Generalized vector database systems, such as pgvector~\cite{pgvector}, VBASE~\cite{vbase}, AnalyticDB~\cite{AnalyticDB},  and DuckDB~\cite{duckdb}, integrate vector data directly within relational databases, making them better suited to address these limitations~\cite{Lucene}.
However, they still face significant challenges, especially the lack of accurate cardinality estimation for vector similarity predicates. Although these systems can leverage existing cost-based query optimizers to generate efficient execution plans~\cite{survey-vector}, their approaches mainly target simple filter queries over a single relation~\cite{vbase, AnalyticDB, singlestore}. In contrast, multi-relation VAQs with complex joins require more precise cardinality estimation~\cite{errors-in-cardinality}, which current systems have not yet fully addressed.
As illustrated in \autoref{fig:motivation}, inaccurate cardinality estimation in vector similarity search can lead to suboptimal query execution plans.
The root cause is that most generalized vector database systems employ heuristic methods, as shown in \autoref{tab:vss_magic_number}. Database systems such as pgvector, VBASE, and DuckDB use simplistic fixed selectivity values for VSS operators regardless of query predicates or data distribution, which often misleads the optimizer and results in significant suboptimal query plans.
\\} 
\begin{table}[t!]
\centering
\caption{
Heuristic fixed selectivity values used by generalized vector database systems for estimating the cardinality of VSS operators. $\mathit{Card}(T)$ is the number of rows in the table T containing vector data.}
\label{tab:vss_magic_number}
\begin{tabular}{l c c}
\hline
\textbf{DBMS}  & \textbf{Selectivity} & \textbf{Estimated Cardinality} \\
\hline
pgvector   & Fixed &$0.333 \times \mathit{Card}(T)$ \\
VBASE      & Fixed &$0.500 \times \mathit{Card}(T)$ \\
DuckDB     & Fixed &$1.000 \times \mathit{Card}(T)$ \\
\hline
\end{tabular}
\vspace{-4mm}
\end{table}
In our empirical analysis, incorporating accurate cardinality estimates for vector similarity search resulted in significantly more efficient query plans. Improvements in join order, join type, and scan strategy led to substantial performance gains by reducing intermediate computation and avoiding unnecessary work. These results underscore the importance of precise cardinality estimation in optimizing query execution for VAQs.
This improvement becomes even more critical as dataset sizes scale~\cite{ce_important_microsoft, cardinality_important}, highlighting the essential role of accurate cardinality estimation in optimizing VAQs within generalized vector database systems~\cite{errors-in-cardinality}.

\section{Vector-augmented SQL analytics}
\label{sec:benchmark}

VAQs extend traditional analytical workloads by integrating relational operations with VSS. While existing ANN benchmarks effectively evaluate vector indexing and search algorithms, they primarily focus on high-recall and high-QPS scenarios over a single relation with limited structured filtering. Consequently, they fail to reflect the complexity of real-world analytical workloads, which often require multi-attribute filtering, relational joins, and fine-grained similarity conditions as part of end-to-end query execution. This gap limits the applicability of such benchmarks for evaluating the performance and functionality of modern vector-aware database systems. 
\change{To address this limitation, we design a benchmark called \textit{Vector-augmented SQL analytics}, which extends the widely used TPC-H ~\cite{tpc-h} and TPC-DS~\cite{tpc-ds_1} benchmarks by incorporating realistic vector data into both datasets and queries.}

\change{
This benchmark is designed to capture the complexity of real-world workloads by enabling systematic evaluation of VAQs in hybrid settings that combine structured relational data with high-dimensional vector representations. We augment the \textit{partsupp} table in TPC-H with two embedding columns (\textit{ps\_image\_embedding} and \textit{ps\_text\_embedding}) and an additional tag column (\textit{ps\_tag}) to simulate multimodal characteristics of parts provided by suppliers. These embeddings are derived from widely used datasets such as DEEP~\cite{deep}, SIFT~\cite{sift}, SimSearchNet++~\cite{simsearchnet}, YFCC~\cite{yfcc,annbenchmark23}, and WIKI~\cite{huggingface-wikipedia}, ensuring diverse and realistic vector characteristics across image and text domains. We further extend the schema by adding a text embedding column (\textit{p\_text\_embedding}) to the \textit{part} table, enabling evaluation of queries that involve multiple vector predicates across relations as well as combinations of vector similarity and attribute-based filtering.}

To construct vector-augmented analytical queries, we extend representative TPC-H queries: Q3, Q5, Q8, Q9, Q10, Q11, Q12, and Q20. These queries were selected to reflect a diverse range of query characteristics, including variations in the number of joins, the join paths between tables, and the presence of complex filtering~\cite{tpch-quantifying}.
We further chose queries that directly use or are well-suited for augmentation with the \textit{partsupp} table.
We preserve the original relational filter conditions to maintain semantic consistency, while adding vector-based predicates to reflect modern hybrid workloads. The resulting queries integrate traditional filters with vector similarity conditions, creating realistic analytical scenarios that jointly stress relational and vector processing components.

\shorten{
A key feature of \textit{Vector-augmented SQL analytics} is the introduction of a \textit{vector distance threshold}, denoted as $\texttt{ps\_embedding <-> \$\{image\_embedding\}} < \$\{D\}$. 
This enables range-based vector similarity search tailored to real-world use cases such as anomaly detection, contextual recommendations, and similarity-based visual filtering. Whereas top-$k$ search always returns a fixed number of results regardless of their actual similarity, it may yield results that are either semantically weak or fail to capture the full set of strongly related vectors. In contrast, vector range queries apply a user-defined similarity threshold $D$ to retrieve only those records that meet a minimum similarity requirement. Furthermore, this threshold condition also provides benchmarking flexibility, allowing one to vary query selectivity and test system performance under different constraints. Unlike top-$k$ search with fixed cardinality, vector range queries return a variable number of results based on the threshold. This makes cardinality estimation significantly more challenging, providing a practical basis for optimizers facing uncertainty in modern hybrid workloads.
}

\change{In addition, we extend our benchmark design to TPC-DS~\cite{tpc-ds_1}, following a similar approach to TPC-H. Specifically, we augment the \textit{item} table with an embedding column (\textit{i\_embedding}) and queries (Q7, Q12, Q19, Q20, Q42, Q72, and Q98) are selected across diverse query classes~\cite{tpc-ds_2} and varying selectivities to evaluate VAQs under more complex analytical scenarios.}

\section{\sysname}
\label{sec:exqutor}

\begin{figure} [tb!]
    \centering
    \includegraphics[width=0.99\linewidth]{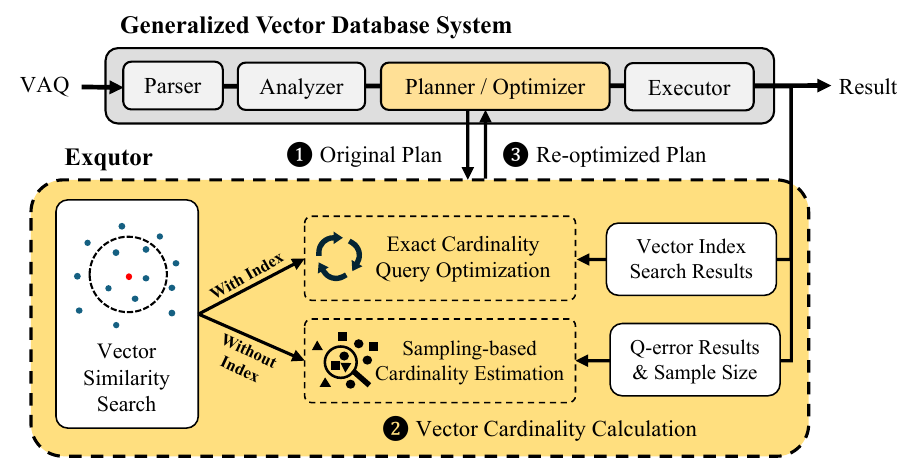}
    \caption{
    Integration of \sysname into a generalized vector database system. When a VAQ is processed, the original query plan is forwarded to \sysname (\ding{202}), which calculates vector cardinality using ECQO or sampling-based cardinality estimation (\ding{203}). The estimated cardinality for vector range search is then returned to the query optimizer, allowing it to generate a more accurate and efficient execution plan (\ding{204}).
    }
    \label{fig:exqutor}
    \vspace{-4mm}
\end{figure}

\shorten{
We present \sysname, a system designed to enhance the query optimization process for VAQs in generalized vector database systems by estimating the cardinality of vector range search. As illustrated in \autoref{fig:exqutor}, \sysname integrates seamlessly into existing query optimizer pipelines.
It supports both ANN and KNN queries and operates during the query planning phase to influence the selection of optimal execution strategies. 
For VAQs with vector indexes, \sysname employs Exact Cardinality Query Optimization (ECQO), which performs lightweight index-based searches and derives accurate cardinality values for vector range queries during query planning (\autoref{sec:ECQO}). For VAQs without index, \sysname uses a sampling-based approach to approximate selectivity (\autoref{sec:sampling}).}




\subsection{Vector Index-based ECQO}
\label{sec:ECQO}

When a VAQ involves a vector similarity predicate and a corresponding index is available, \sysname applies a strategy called ECQO. The key idea behind ECQO is to execute a lightweight vector index search during query planning to compute the exact number of vectors among the retrieved candidates that satisfy the similarity threshold. 
The resulting cardinality is incorporated into the optimizer’s cost model, guiding the selection of both join ordering and scan strategies for more efficient plan generation. By replacing heuristic assumptions with precise estimates, ECQO helps the optimizer construct execution plans.

\shorten{
Many vector database systems support ANN indexes such as HNSW and IVF, which are primarily designed to accelerate similarity search during query execution. ECQO repurposes these index structures during the planning phase by issuing a range query over the index using the actual query vector and user-defined similarity threshold. Since ANN indexes such as HNSW are designed to limit the search space using hierarchical or graph-based traversal strategies, even when invoked during planning, the index search completes quickly and introduces negligible overhead.
Although ANN indexes are approximate by design, they can exhibit deterministic behavior with fixed configurations. For instance, in the case of HNSW, the result of a search remains deterministic as long as the index graph, entry point, search parameters (e.g., \(ef\_search\)), and query vector remain unchanged. This property allows ECQO to treat the cardinality obtained during query planning as an exact input to the optimizer. 
}

In addition to its accuracy, ECQO improves efficiency by reusing computation across planning and execution. Since the index search used during planning retrieves the same candidate vectors as those needed at execution time, \sysname can cache the results and eliminate redundant searches. As a result, the ANN vector search only needs to be performed once per query, effectively reducing execution latency and avoiding unnecessary computation. Collectively, the accuracy, determinism, and low overhead of ECQO make it a practical and effective mechanism for optimizing vector-aware queries in systems that support vector indexing.

\mypar{Implementation in generalized vector database systems}
\shorten{
We integrated ECQO into multiple systems based on pgvector, VBASE, and DuckDB by extending their query optimizers to incorporate exact cardinality estimates obtained from vector indexes.
In pgvector and VBASE, we integrated ECQO by extending the planner's hook to identify VAQs with vector range predicates and trigger index-based similarity searches during query planning. 
To minimize redundant computation across query stages, the candidate set retrieved during planning is cached and reused during execution. Given that ANN index searches produce deterministic results under fixed parameters, this reuse ensures consistency while reducing overhead. In DuckDB, we implemented ECQO by modifying the logical optimizer rules responsible for cardinality estimation. 
DuckDB's in-process architecture enables efficient propagation of this estimate to downstream components of the optimizer, including join reordering and scan strategy selection. 

}



\subsection{Sampling-based Cardinality Estimation without Vector Index}
\label{sec:sampling}
When a VAQ lacks a vector index, the query optimizer must rely on either an index over structured attributes or perform a full sequential scan. In the case of a sequential scan, evaluating the similarity predicate requires computing distances between the query vector and all vectors in the dataset. This exhaustive KNN search is highly expensive, making it unsuitable for direct execution during query planning, unlike the approach used in ECQO. 
To address this, \sysname adopts a sampling-based cardinality estimation approach specifically for KNN queries, where it approximates the number of qualifying tuples by evaluating similarity over a small subset of the data. This enables the optimizer to obtain meaningful cardinality estimates at a fraction of the cost of a full scan, making sampling a practical alternative for query planning in the absence of vector indexes. Similar to ECQO, the estimated cardinality is integrated into the optimizer's cost model, allowing it to select execution plans that better reflect the selectivity of the vector range predicate.

To determine an appropriate sample size, \sysname uses a statistical formula derived from classical sampling theory~\cite{sample_size}. The required number of samples $N$ is computed as:

\begin{equation}
\small
    N = \left\lceil \frac{z^2 \cdot \hat{P} \cdot (1 - \hat{P})}{e^2} \right\rceil
    \label{eq:sampling_size_equation}
\end{equation}
\begin{description} 
    \item[$z$] critical value corresponding to the desired confidence level (e.g., $z = 1.96$ for 95\% confidence).
    \item[$\hat{P}$] estimated proportion of data points expected to fall within the similarity threshold.
    \item[$e$] desired margin of error (e.g., $e = 0.05$ for 5\% error).
\end{description}

\mypar{Adaptive sampling size adjustment}
While fixed sample sizes provide statistical guarantees, they may not be equally effective across datasets with varying distributions or dimensionalities. In high-dimensional or skewed datasets, a fixed sample size may be unnecessarily large, resulting in wasted resources, or too small, leading to inaccurate estimates. To address this, \sysname introduces an adaptive sampling mechanism that dynamically adjusts the sample size based on estimation accuracy observed after query execution. This mechanism aims to balance estimation precision with computational cost, adapting to the workload characteristics.

\sysname employs a momentum-based adjustment algorithm combined with a learning rate scheduler to adapt the sampling size over time. Momentum smooths fluctuations in adjustment, preventing instability, while the learning rate scheduler gradually reduces update magnitude to ensure convergence. The adjustment is guided by the Q-error~\cite{learned, deepdb, selectivity}, which measures the deviation between the estimated and true cardinality:
\begin{equation}
    \text{Q-error} = \max\left( 
    \frac{\text{Card}_\text{esti}}{\text{Card}_\text{true}},\ 
    \frac{\text{Card}_\text{true}}{\text{Card}_\text{esti}} 
    \right)
\end{equation}

Using this metric, \sysname tracks recent estimation accuracy and updates the sample size according to the following rule:
\begin{equation}
    \delta = \alpha \cdot (\text{Q-error} - \beta) - (100 - \alpha) \cdot \text{sampling\_ratio}
\end{equation}
\begin{equation}
    V_t = m \cdot V_{t-1} + \eta_t \cdot \delta
\end{equation}
\begin{equation}
    \textit{sampling\_size}_{t+1} = \textit{sampling\_size}_{t} + V_t
\end{equation}

Here, $\delta$ is the adjustment factor computed from estimation error and the current sampling ratio, which determines the direction and magnitude of sample size updates. $V_t$ is the momentum term at iteration $t$, $m$ is the momentum coefficient, and $\eta_t$ is the learning rate. $\alpha$ balances the contribution between Q-error and the sampling ratio, and $\beta$ is a tunable threshold representing acceptable Q-error.




The learning rate is decayed at each iteration using:
\begin{equation}
    \eta_{t+1} = \gamma \cdot \eta_t
\end{equation}
where $\gamma$ is the decay factor ($0 < \gamma < 1$) that progressively reduces the adjustment magnitude.

This adaptive mechanism enables \sysname to respond effectively to changing query workloads and data characteristics. When estimation remains accurate with low Q-error, the sample size is reduced to save computation. Conversely, higher Q-error triggers an increase in sample size to restore accuracy. 
This feedback-driven adaptation ensures that sampling remains both efficient and reliable over time.

By combining statistical sampling theory with adaptive learning techniques, \sysname delivers a practical and robust solution for cardinality estimation in vector similarity queries without index support. This method is particularly effective for exploratory and analytical queries on large datasets, where full scans are infeasible and traditional estimates are insufficient.

\mypar{Implementation in generalized vector database systems}
\shorten{
We implemented sampling-based cardinality estimation in pgvector by extending the query optimizer to support dynamic sample size adjustment during planning. When a VAQ with a vector range predicate lacks index support, the optimizer invokes a sampling routine that evaluates the similarity predicate over a representative subset of the data, rather than performing a full KNN scan. 
To support adaptive sampling, we extended the optimizer to track estimation accuracy using the Q-error metric. After each query, the system compares the estimated cardinality with the actual value observed during execution and uses the resulting Q-error as feedback to adjust the sample size for future queries, expanding it when accuracy is insufficient and shrinking it when estimates remain stable. Additionally, the optimizer maintains separate sample size states for each table, allowing it to adapt to the specific distributional characteristics of different datasets. 

}





\section{EVALUATION}
\label{sec:eval}





We evaluate the performance of \sysname by integrating it into pgvector, VBASE, and DuckDB, demonstrating its generality and pluggability across fundamentally different database system architectures. Our evaluation focuses on improvements in cardinality estimation accuracy, query execution performance, and scalability across varying data and workload characteristics. In particular, we show how \sysname enhances cardinality estimation for vector range predicates in Vector-augmented Analytical Queries, enabling more effective query optimization and yielding substantial execution-time improvements.


\change{
To assess the impact of ECQO and sampling-based cardinality estimation, we design experiments for two types of VAQs. 
First, we evaluate the benefit of executing lightweight vector index searches during query planning for VAQs (\autoref{eval:index}). 
Second, we analyze how sampling-based cardinality estimation improves VAQs without index support (\autoref{eval:sampling}). 
We then extend the evaluation to diverse workloads, including multi-vector queries, correlation-aware queries, and TPC-DS (\autoref{sec:div_workloads}), 
and conclude with an in-depth analysis of \sysname covering overhead, scalability (\autoref{sec:discussion}), and limitations (\autoref{sec:limitations}). 
Our key findings are summarized as follows:
}



\begin{itemize}[noitemsep,topsep=3pt,parsep=0pt,partopsep=0pt,leftmargin=10pt]
    \item For VAQs utilizing vector indexes, \sysname significantly improves query execution performance. In PostgreSQL-based systems, pgvector and VBASE, ECQO achieves speedups ranging from 1.01{\small$\times$} to four orders of magnitude. In DuckDB, ECQO yields speedups from 1.5{\small$\times$} to 37.2{\small$\times$}.
    \item For VAQs without vector indexes, sampling-based cardinality estimation improves performance from 1.2{\small$\times$} to 3.2{\small$\times$}. The adaptive strategy converges to dataset-specific sample sizes, balancing estimation accuracy and planning overhead across diverse vector distributions.
    \item \sysname consistently improves cardinality estimation accuracy under diverse query conditions, including different selectivities, data scales, and vector characteristics, while incurring negligible overhead, enabling robust and efficient query planning.
\end{itemize}

\mypar{Datasets and VAQs} 
\change{We conduct experiments on TPC-H and TPC-DS based \benchmarkname using widely used vector datasets: DEEP (96 dimensions) \cite{deep}, SIFT (128 dimensions) \cite{sift}, SimSearchNet++ (256 dimensions) \cite{simsearchnet}.
These datasets not only vary in embedding dimensionality but also cover diverse data distributions, including skewed and normal \cite{data-distribution}, which are commonly observed in real-world vector workloads.
These datasets represent realistic use cases in multimedia retrieval and allow us to investigate how vector dimensionality affects query performance. For each dataset, we construct VAQs\footref{fn:github} that perform filtering alongside vector similarity search, reflecting practical analytics scenarios.}

To ensure consistency across experiments, we configure range thresholds in VAQs such that the expected number of matches is controlled. For index-based VAQs, the range threshold is tuned to return 200 vectors. For sampling-based VAQs, the threshold is set to retrieve 1\% of total rows. This setting reflects typical usage where the user seeks only closely related vectors and avoids biasing the experiments toward either \sysname or the baselines.

\mypar{System setup}  
We conduct our experiments using pgvector, VBASE, and DuckDB with \texttt{DuckDB-VSS}. The system is equipped with Intel Xeon Gold 6530 configured with 128 vCPUs and 1.0 TB of RAM. 

Each experiment begins with a warm-up execution, which is excluded from the reported results to eliminate caching effects. We then report the trimmed mean of execution times over ten runs, removing the lowest and highest runtimes. 
We run with default settings, where PostgreSQL used 8 for \(max\_worker\_processes\) and DuckDB used 128 for \( worker\_threads\).



\mypar{Indexing and sampling parameter configuration}
For VAQs involving vector indexes, we use HNSW \cite{hnsw} as the underlying ANN structure in both \sysname and the baseline systems. We configure HNSW with the same vector index parameters (\(M = 16\), \(ef\_construction = 200\), \(ef\_search = 400\)). 


For sampling-based cardinality estimation, we initially compute the number of samples \(N\) using the sample size  formula (\autoref{eq:sampling_size_equation}) for sample size estimation \cite{sample_size}, given a 95\% confidence level (\(z = 1.96\)), a proportion estimate \(\hat{P} = 0.5\), and a 5\% margin of error (\(e = 0.05\)). Applying the formula yields a fixed sample size of \(N = 385\).

For adaptive sampling, we extend the optimizer with momentum-based feedback control. Parameter values are selected based on prior work on adaptive query estimation \cite{momentum, selectivity}: we set the momentum coefficient \(m = 0.9\), initial learning rate \(\eta_0 = 0.1\), weighting factor \(\alpha = 50\), and target Q-error \(\beta = 1.5\). These values balance Q-error minimization and sample size stability. The learning rate decay factor \(\gamma = 0.99\) gradually reduces adjustment magnitude to ensure convergence. Sample size updates are triggered every 50 queries. 
\vspace{-1mm}

\subsection{Vector Index-based Exact Cardinality Query Optimization}
\label{eval:index}



    


In this section, we evaluate the performance of \sysname when executing VAQs with a vector index using an ANN search, specifically with HNSW~\cite{hnsw}. The experiments compare two configurations: (i) baseline generalized vector database systems, and (ii) execution with ECQO enabled. We evaluate \sysname integrated into pgvector, VBASE, and DuckDB, analyzing its effectiveness in optimizing query plan by injecting exact cardinality and reducing redundant computations through lightweight index-based vector search.

\begin{figure}[t!]
    \centering
    \begin{subfigure}[b]{0.99\linewidth}
        \centering
        \includegraphics[width=\linewidth]{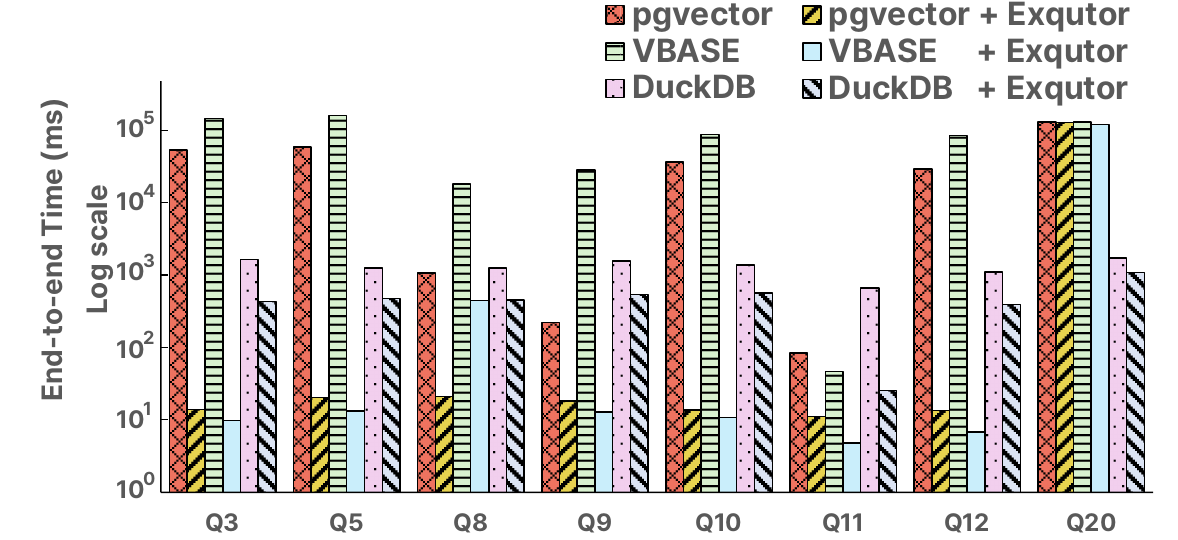}
        \caption{DEEP}
        \label{fig:e2e_deep}
    \end{subfigure}
    \begin{subfigure}[b]{0.99\linewidth}
        \centering
        \includegraphics[width=\linewidth]{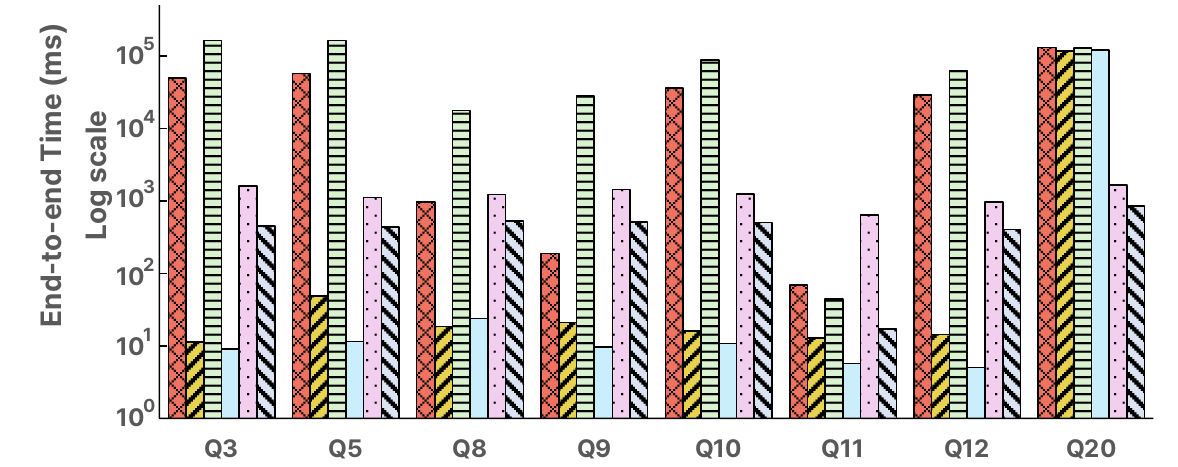}
        \caption{SIFT}
        \label{fig:e2e_sift}
    \end{subfigure}
    \begin{subfigure}[b]{0.99\linewidth}
        \centering
        \includegraphics[width=\linewidth]{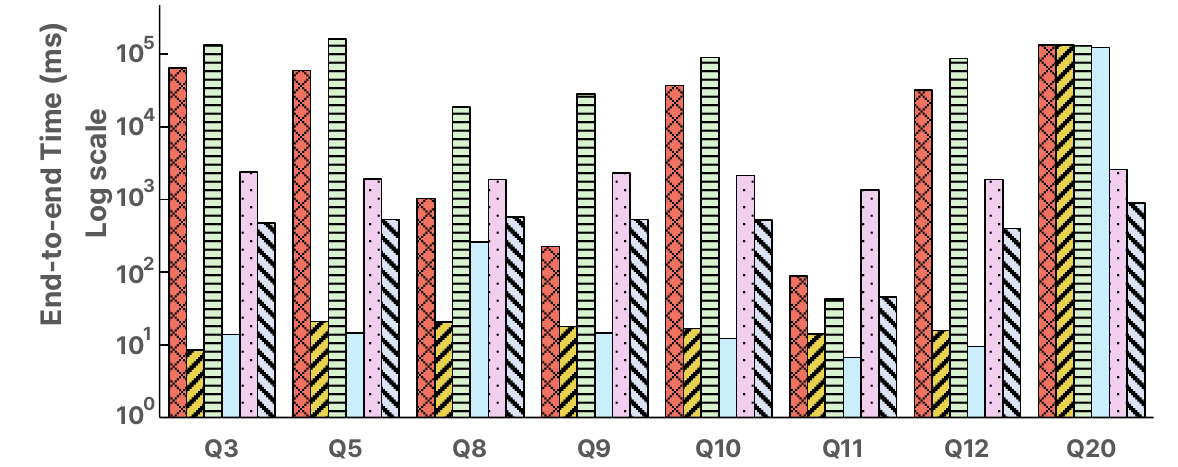}
        \caption{SimSearchNet++}
        \label{fig:e2e_fb}
    \end{subfigure}
    
    \caption{Query execution time for TPC-H VAQs with a vector index using three different datasets (SF100). Each subfigure compares query latency with and without \sysname integration in pgvector, VBASE and DuckDB.}
    \label{fig:ECQO_excution_time}
    \vspace{-6mm}
\end{figure}

\mypar{Performance gains from ECQO}
\autoref{fig:ECQO_excution_time} shows end-to-end query execution times for TPC-H based \benchmarkname using three datasets. Each figure compares six system configurations, including pgvector, VBASE, and DuckDB, with and without \sysname integration. Across all datasets and systems, ECQO consistently improves query performance by enabling accurate cardinality estimation during planning.

The most substantial performance gains are observed in pgvector and VBASE. When \sysname is integrated into pgvector, it achieves up to three orders of magnitude speedup over the baseline. Similarly, applying \sysname to VBASE yields speedups of up to four orders of magnitude by enabling accurate cardinality estimation for vector range search. 
These improvements stem from two key mechanisms. First, ECQO performs a lightweight vector index probe during planning using the HNSW structure, which returns the exact number of qualifying tuples. This cardinality is incorporated into the cost model, improving scan method selection and join ordering decisions. In all evaluated queries, the \textit{partsupp} table, which contains the vector predicate, is selected as the first input in the join plan, enabling early filtering and reducing intermediate result sizes. 
Second, \sysname caches the retrieved index results during planning and reuses them during execution, avoiding redundant similarity computations.

In DuckDB, which already features a highly optimized execution engine for analytic queries, ECQO still yields measurable improvements. While the baseline query plans in DuckDB are already efficient, ECQO enhances query planning by exposing the true selectivity of vector range search. This results in improved join strategies and execution time reductions for VAQs.

\shorten{
\mypar{Cardinality estimation and query plan optimization}  
In PostgreSQL-based systems, pgvector and VBASE, the query optimizer lacks native support for vector-aware statistics. Consequently, cardinality estimates for vector predicates default to fixed or arbitrary values, regardless of data size or threshold selectivity. This leads to poor plan choices such as unnecessary hash joins or late application of selective filters. With \sysname, the optimizer receives precise cardinality derived from the actual vector index search. This improves plan quality by enabling the planner to choose nested loop joins when beneficial and to push down vector filters early. Sequential scans are often replaced with index scans, reducing I/O and computation.

DuckDB exhibits similar issues, as it also lacks built-in selectivity estimation for vector data. Without ECQO, its optimizer must assume worst-case cardinality for similarity predicates. By integrating ECQO through \sysname, DuckDB can leverage index-based estimates during planning, producing more accurate join orders and improving execution. While the relative gains are smaller than in PostgreSQL, the improvement is still consistent across datasets.

Interestingly, our evaluation reveals that pgvector and VBASE with \sysname can outperform baseline DuckDB on certain queries. This is due to PostgreSQL's ability to benefit more from ECQO’s index scan enablement, which offsets its slower executor. These results show that optimizer enhancements like ECQO can shift performance bottlenecks, making even traditionally slower systems competitive with modern execution engines.
}

\mypar{Impact of vector dimensionality and query characteristics}
With ECQO, the availability of accurate cardinality estimates for vector range search allows the optimizer to avoid unnecessary operations such as full scans and hash joins, significantly reducing the cost of non-vector operators. As a result, HNSW-based vector search emerges as the dominant component of query execution time. Because the cost of HNSW search increases with vector dimensionality due to more expensive distance computations, we observe a corresponding rise in overall query time as the dimensionality grows.

Despite these improvements, PostgreSQL’s cost model remains insufficiently equipped to accurately reflect the performance characteristics of ANN-based vector indexes. In pgvector, although the index leverages PostgreSQL’s internal buffer pool, ANN indexes typically incur high space amplification~\cite{ann_index_large_size}, often exceeding the size of the base table. \change{At the same time, structures like HNSW achieve $O(\log \mathit{Card}(T))$ sublinear search times~\cite{hnsw}, which the current cost model fails to capture. As a result, the optimizer may overestimate the cost of an HNSW index scan and fail to select it, even in cases where it would be the better access method \cite{overestimation_problem}.}
As a result, the system significantly overestimates the cost of index scans. VBASE, on the other hand, decouples ANN index access from the primary buffer pool and manages it with an independent memory structure, making precise cost estimation even more difficult. These limitations can still result in suboptimal plan choices, indicating that further refinements to the underlying cost model are needed to fully capitalize on accurate cardinality estimates.

Query characteristics also influence ECQO’s effectiveness. Most evaluated queries benefit from better join ordering and early application of vector filters. However, in queries like Q20, where the dominant cost arises from a full scan of the \textit{lineitem} table, ECQO’s impact is limited. While ECQO improves join ordering, the total benefit is limited by the large fixed cost of scanning unrelated data. This suggests that ECQO is most effective when vector predicates contribute significantly to overall selectivity.

\subsection{Sampling-based Cardinality Estimation}
\label{eval:sampling}

\begin{figure}[t!]
    \centering
    \includegraphics[width=0.99\linewidth, trim={0pt 10pt 0pt 15pt},clip]{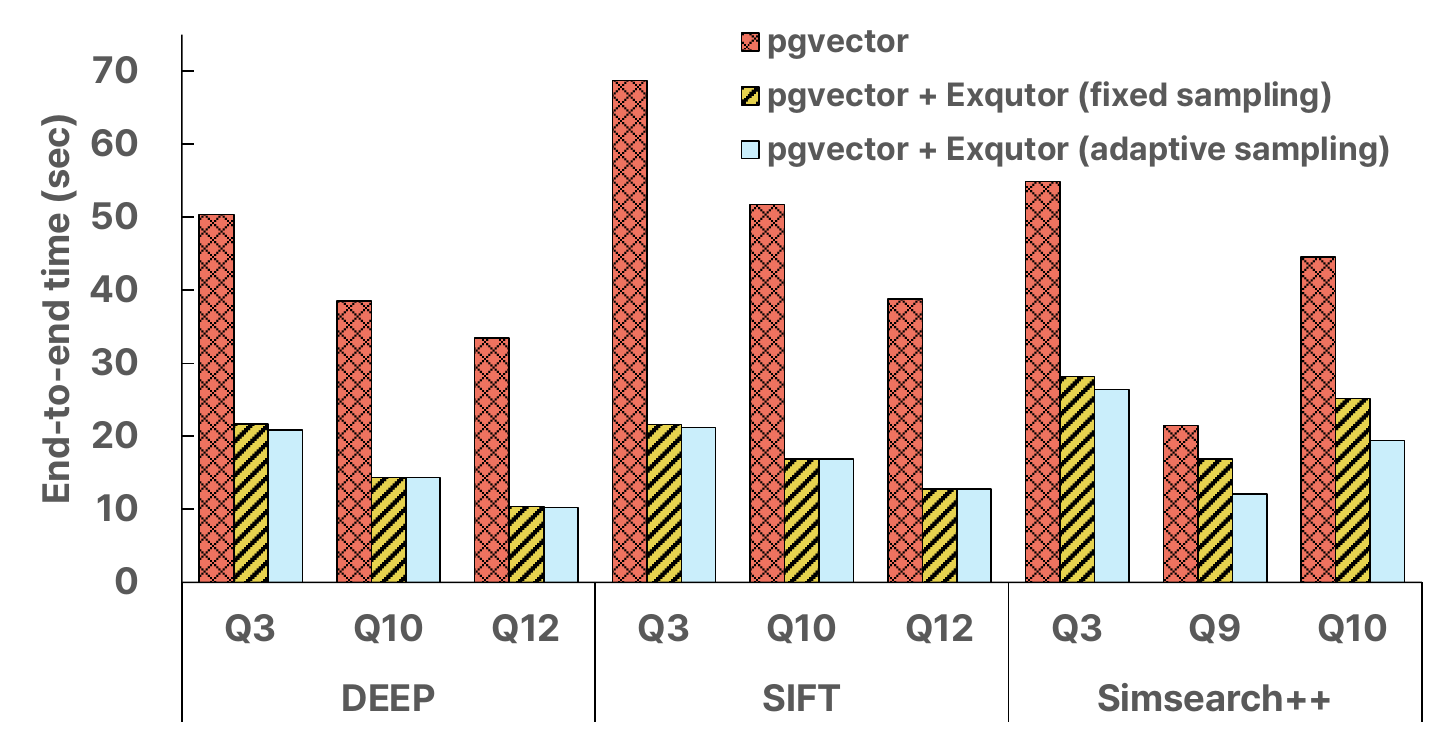}
    \caption{Query execution time on pgvector for TPC-H VAQs without vector index (SF100). The fixed sample size uses a constant sample size of  \(N = 385\), whereas the adaptive sampling strategy dynamically adjusts the sample size based on Q-error.}
    \label{fig:pg_sampling}
    \vspace{-6mm}
\end{figure}

In this section, we evaluate the performance of \sysname applied to TPC-H VAQs that perform KNN searches without vector indexes, where cardinality estimation is handled via sampling.
We compare three configurations: (i) the baseline pgvector execution without sampling, which uses default selectivity estimates, (ii) \sysname with a fixed sample size derived from statistical confidence bounds, and (iii) \sysname with adaptive sampling that dynamically adjusts the sample size based on query feedback and dataset properties. For consistency, we focus on a subset of VAQs where the optimizer selects a sequential scan on the \textit{partsupp} table which means KNN search forms a dominant component of execution cost. These queries are representative of realistic cases in vector analytics without index support.


\mypar{Performance gains from sampling-based cardinality estimation}
\autoref{fig:pg_sampling} shows that both fixed and adaptive sampling significantly improve execution time compared to the baseline. Fixed sampling achieves speedups from 1.2{\small$\times$} to 3.2{\small$\times$}, demonstrating that even a small, uniform sample provides better cardinality feedback. However, fixed-size sampling does not consider data distribution or vector dimensionality. For example, when working with dense clusters or high-dimensional embeddings, the fixed sample may misrepresent selectivity, leading to misestimation and suboptimal plans.

Adaptive sampling overcomes this limitation by modifying the sample size based on query feedback. 
It tracks Q-error over time and adjusts the number of sampled rows accordingly. 
When the error is high, indicating that the estimate diverges from observed cardinality, the sample size is increased to enhance accuracy. Conversely, when estimates stabilize, the sample size is reduced to conserve computation. 
Adaptive sampling delivers up to 1.4{\small$\times$} speedup over fixed sampling, consistently outperforming fixed sampling across datasets with varying distributional properties. The ability to react to query conditions makes adaptive sampling especially effective for dynamic workloads and heterogeneous data.

\begin{figure}[t!]
    \centering
    \includegraphics[width=0.99\linewidth, trim={0pt 2pt 0pt 8pt},clip]{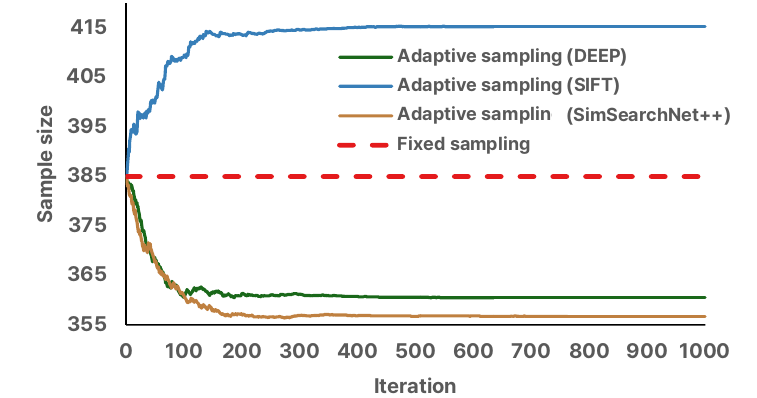}
    \caption{Convergence of adaptive sample size over time on the DEEP, SIFT, and SimSearchNet++ (SF100).
The plot illustrates how \sysname adaptively adjusts sample sizes for VAQs compared against fixed sampling.}
    \label{fig:adaptive_sample_size}
    \vspace{-6.5mm}
\end{figure}

\mypar{Effect of adaptive sampling}  
To understand how adaptive sampling evolves over time, \autoref{fig:adaptive_sample_size} illustrates how \sysname adjusts the sample size in response to Q-error. Initially, the system starts with a statistically determined sample size using \autoref{eq:sampling_size_equation} and evaluates estimation accuracy after each query. When the Q-error exceeds a predefined threshold, \sysname increases the sample size using momentum-based updates, which smooth out fluctuations and promote stable convergence. As updates continue, the learning rate decays gradually, leading to smaller adjustments over time. This feedback loop enables the system to maintain estimation accuracy while minimizing unnecessary computation.

This behavior demonstrates that \sysname effectively balances estimation accuracy and planning efficiency. The sample size trajectory varies depending on the dataset: for DEEP and SimSearchNet++, the sample size decreases over time as Q-error stabilizes, allowing the system to reduce planning cost without loss of accuracy. In contrast, for SIFT, the sample size increases to satisfy higher estimation demands due to its more complex distribution. Ultimately, \sysname converges to a dataset-specific equilibrium that reflects the selectivity patterns and estimation difficulty of each workload.

\mypar{Accuracy and optimizer impact}
\shorten{
The effectiveness of sampling-based estimation directly translates to improved query plans. Without sampling, pgvector relies on fixed default cardinality values, which often result in suboptimal join ordering and delayed application of vector predicates. For example, due to underestimated selectivity, the optimizer may place the join with the \textit{partsupp} table later in the execution plan, which results in unnecessarily large intermediate results and degraded performance. In contrast, sampling allows \sysname to provide more accurate cardinality estimates, enabling the optimizer to apply vector filters earlier and improve the plan across multiple aspects. Hash joins are often replaced with nested loop joins that better exploit early filtering, and scan strategies also improve, with the optimizer selecting index-based access over full sequential scans.


Overall, sampling-based cardinality estimation is a lightweight yet powerful technique for improving query optimization in VAQs without index support. Fixed sampling provides a simple, statistically grounded baseline that already improves execution time. Adaptive sampling further enhances this by learning from query performance and dynamically tuning sampling effort. 
Together, these methods allow \sysname to apply selectivity-aware optimization even when vector indexes are unavailable, while also ensuring stable performance under shifting workloads.
}
\change{
\begin{figure*}[t]
    \centering
    \begin{minipage}[t]{0.32\textwidth}
    \includegraphics[width=\linewidth]{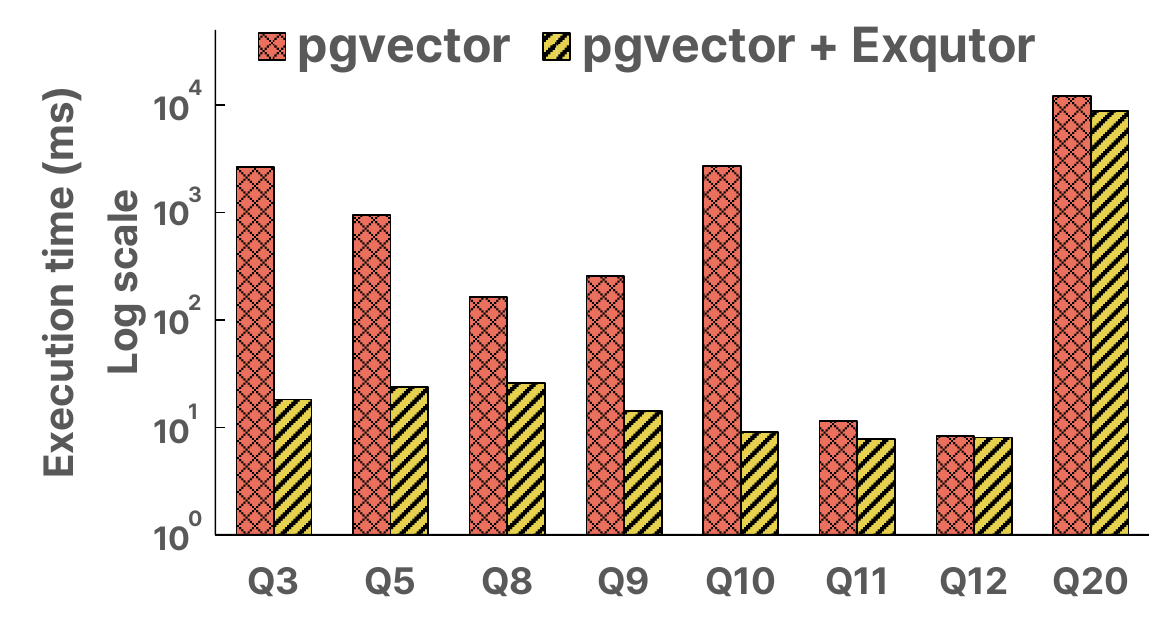}
    \caption{Query execution time for TPC-H VAQs on the \texttt{partsupp} table with YFCC and tag filtering (SF10).}
    \label{fig:yfcc}
    \end{minipage}\hfill%
    \begin{minipage}[t]{0.32\textwidth}
    \includegraphics[width=\linewidth]{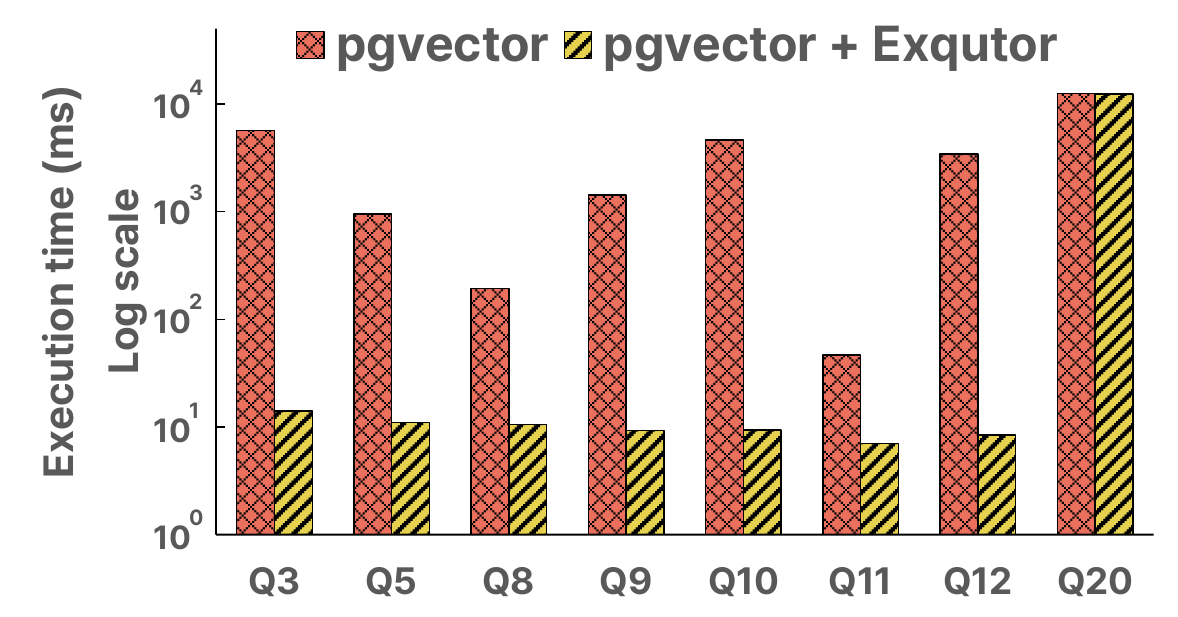}
    \caption{Query execution time for TPC-H VAQs on the \texttt{partsupp} table with DEEP and WIKI (SF10).}
    \label{fig:multi-1-table}
    \end{minipage}\hfill%
    \begin{minipage}[t]{0.32\textwidth}
    \includegraphics[width=\linewidth]{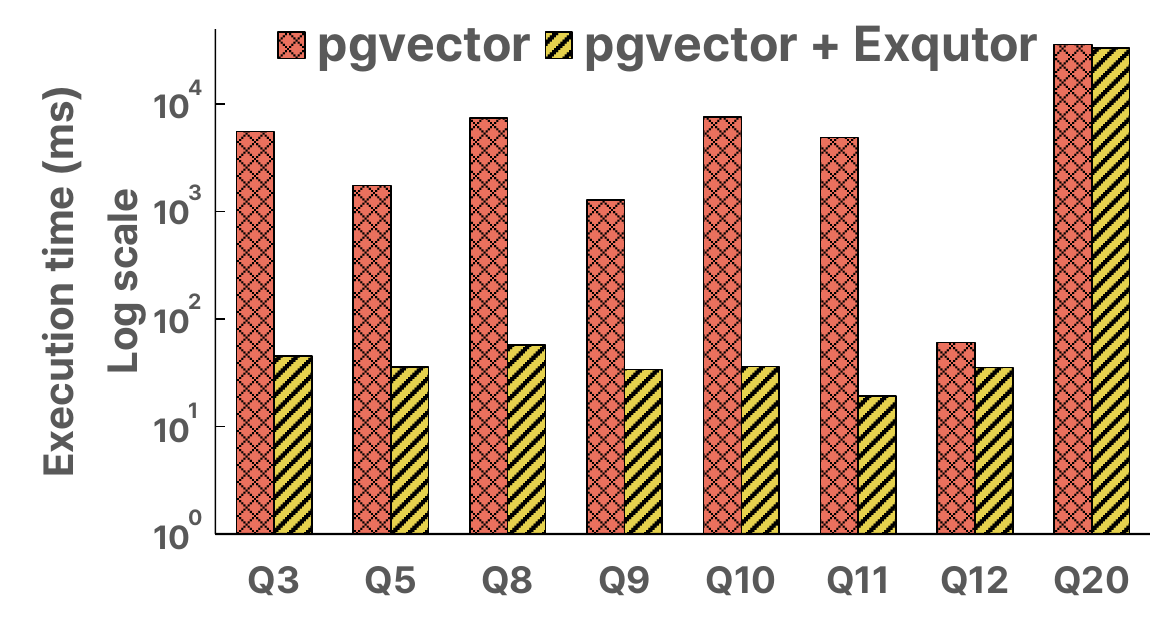}
    \caption{Query execution time for TPC-H VAQs on the \texttt{partsupp} table with DEEP and the \texttt{part} table with WIKI (SF10).}
    \label{fig:multi-2-tables}
    \end{minipage}
    \vspace{-6mm}
\end{figure*}
}

\begin{figure} [t!]
    \centering
    \includegraphics[width=\linewidth, trim={0pt 9.5pt 0pt 13pt},clip]{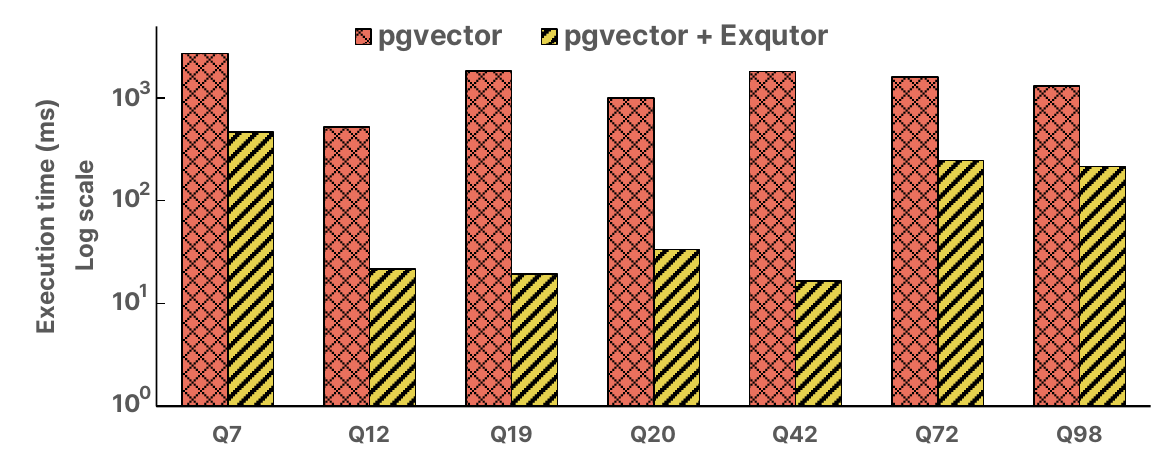}
    \caption{Query execution time for TPC-DS VAQs with vector indexes on the DEEP using pgvector and \sysname (SF10).}
    \label{fig:tpc-ds}
    \vspace{-7mm}
\end{figure}

\change{
\subsection{Performance on Diverse Workloads}
\label{sec:div_workloads}
To further validate the generality of \sysname, we evaluate it on more diverse workloads. 
We extend TPC-H with multi-vector and correlation-aware VAQs that combine vector similarity with tag filtering, and further assess its effectiveness on TPC-DS based \benchmarkname. 
We utilize three widely used embedding datasets, DEEP (96 dimensions)~\cite{deep}, YFCC (192 dimensions)~\cite{yfcc,annbenchmark23} and WIKI (768 dimensions)~\cite{huggingface-wikipedia}.
}
\change{
\mypar{VAQs with correlation}
We evaluate \sysname on VAQs with correlation that combine vector search and tag-based filtering using the YFCC. 
In this setting, the associated tags are stored in \texttt{ps\_tag}, and VAQs are extended with conditions requiring the presence of specific tags. 
As shown in \autoref{fig:yfcc}, query execution times improve significantly, achieving up to 301.5{\small$\times$} speedup. 
\sysname provides accurate cardinality estimates for nodes that jointly apply vector predicates and tag filters, enabling the optimizer to generate more efficient plans.}

\change{
\mypar{Multi-vector VAQs}
We further evaluate \sysname on multi-vector query workloads, where embeddings from multiple sources are integrated into analytical queries. 
As shown in \autoref{fig:multi-1-table}, when both DEEP and WIKI datasets are stored in the \texttt{partsupp} table, we observe substantial performance improvements, with query execution times accelerated by factors ranging from 1.07{\small$\times$} to 479.4{\small$\times$}. 
\autoref{fig:multi-2-tables} illustrates the scenario where DEEP embeddings are stored in \texttt{partsupp} while WIKI embeddings are stored in the \texttt{part} table. 
Even in this more complex join setting, we still observe speedups from 1.07{\small$\times$} to 254{\small$\times$}. 
These results confirm that \sysname effectively adapts to queries involving multiple vector columns across relations, consistently optimizing execution plans and achieving significant gains.
}

\change{
\mypar{Evaluation on TPC-DS}
To further validate the effectiveness of \sysname on diverse workloads, we conducted experiments on the TPC-DS based \benchmarkname, a widely used benchmark that provides rich and complex query templates.
As shown in \autoref{fig:tpc-ds}, the results demonstrate consistent performance improvements, with query execution times achieving speedups of up to 109.6{\small$\times$}.
Moreover, we observed query plan transformations similar to those identified in TPC-H based \benchmarkname, confirming that \sysname effectively adapts to workload diversity in realistic decision-support scenarios.
}

\subsection{Discussion}
\label{sec:discussion}

    

\begin{figure}[t!]
    \centering
    \begin{subfigure}[b]{0.48\linewidth}
        \centering
        \includegraphics[width=\linewidth]{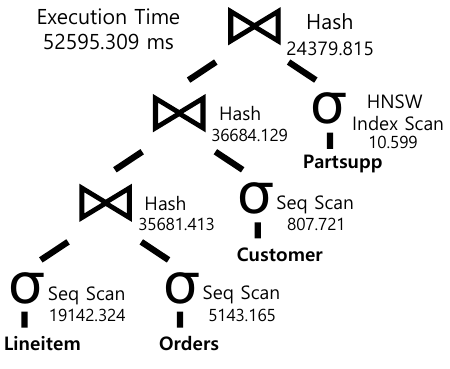}
        \caption{pgvector}
        \label{fig:qa_a}
    \end{subfigure}
    \hfill
    \begin{subfigure}[b]{0.48\linewidth}
        \centering
        \includegraphics[width=\linewidth]{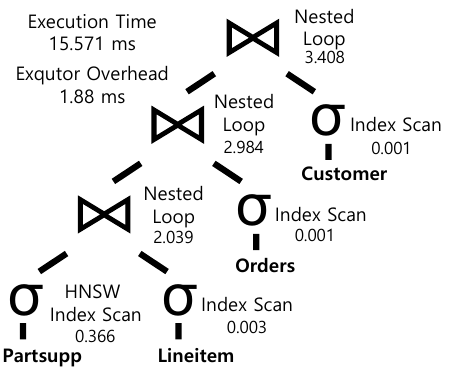}
        \caption{pgvector + \sysname}
        \label{fig:qa_b}
    \end{subfigure}
    \caption{Query plan comparison for Q3 on the DEEP dataset (SF100). (a) shows the execution plan using pgvector, and (b) shows the optimized plan with \sysname, where ECQO enables accurate cardinality estimation using HNSW index probing during planning. Both plans display individual operator nodes along with their execution times, reported in milliseconds.} 
    \label{fig:query_time_analysis}
    \vspace{-6mm}
\end{figure}

\mypar{Query time analysis}
\label{sec: querytime}
\autoref{fig:query_time_analysis} illustrates the execution plans for Q3 on the DEEP dataset before and after applying \sysname with ECQO on pgvector. The baseline pgvector plan relies heavily on full table scans and parallel hash joins, resulting in high scan and join costs, with total query time exceeding 52 seconds. In contrast, the optimized plan with \sysname eliminates expensive join operations and replaces sequential scans with selective index scans, enabled by accurate cardinality feedback from ECQO. 
Notably, the HNSW-based vector range search is executed once during query planning and the result is reused during execution. This reuse reduces the runtime cost of HNSW index scan from 10.6 ms to only 0.366 ms. In addition, ECQO itself contributes only 6.82 ms for cardinality estimation and 1.88 ms as integration overhead. As a result, the total query execution time drops dramatically to 15.571 milliseconds. 
This confirms that exact cardinality injection via ECQO enables the optimizer to generate execution plans that are highly efficient, eliminating unnecessary operations and prioritizing selective access paths. \change{Beyond this single query example, the most common improvements come from join reordering, changes in join methods (e.g., hash joins replaced with nested-loop joins), and scan method transitions from full or bitmap-index scans to index-based access. These patterns consistently appear across different workloads such as TPC-DS, confirming that the benefits of Exqutor generalize beyond a single benchmark. Further illustrations of query execution improvements are provided in supplementary figures\footnote{\url{https://github.com/BDAI-Research/Exqutor/tree/main/query_plans}}.}

\begin{figure}[t!]
    \centering
    \includegraphics[width=\linewidth, trim={0pt 2pt 0pt 10pt},clip]{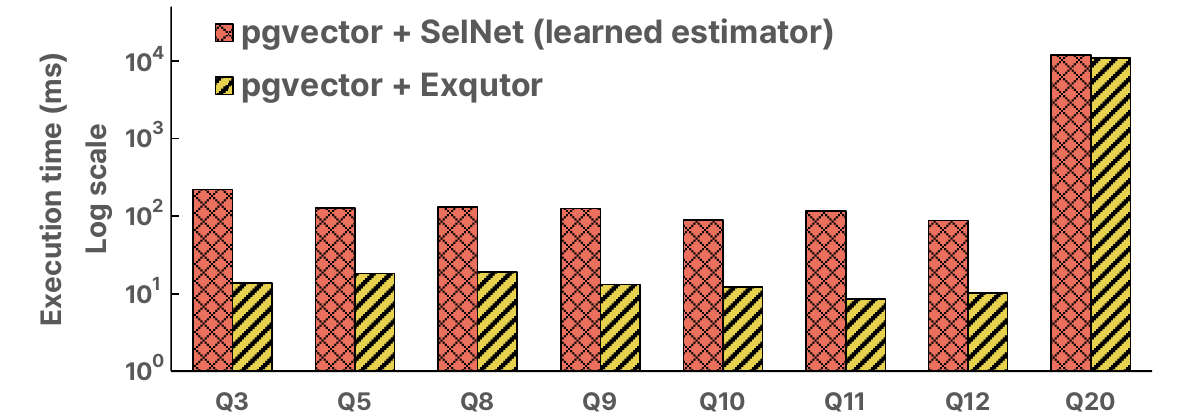}
    \caption{Query execution time of SelNet (learned estimator) and \sysname for TPC-H VAQs with vector indexes on the DEEP dataset (SF10). 
    }
    \vspace{-7mm}
    \label{fig:selnet}
\end{figure}

\change{
\mypar{Comparison with learned cardinality estimator}
\autoref{fig:selnet} compares \sysname with SelNet \cite{selnet}, a learned selectivity estimator. 
\sysname achieves speedups up to 16.1{\small$\times$} speedup over SelNet. 
SelNet requires 77\,ms for a single-query cardinality estimation and depends on offline training, which becomes increasingly costly with the dataset size.
In addition, it relies on external model management, introducing further overhead and complexity.
When compared with the sampling-based approach, \sysname achieves an average Q-error of 1.69, while SelNet yields a higher Q-error of 5.53. 
These results highlight the advantages of \sysname in delivering accurate cardinality estimates with lightweight overhead, ensuring both efficiency and robustness in query optimization.}

\mypar{Cardinality estimation accuracy}
As previously discussed in \autoref{eval:index}, PostgreSQL lacks statistical summaries for vector data, causing it to assign arbitrarily large fixed values for estimated cardinalities. This fixed estimation, irrespective of the given predicate, leads to substantial errors across all datasets.
With static sampling, \sysname significantly reduces estimation errors, achieving Q-errors ranging from 1.04 to 1.57. Adaptive sampling further improves accuracy by adapting the sample size based on Q-error feedback, achieving a lower Q-error range of 1.02 to 1.19, closely aligning with the true cardinality.
\autoref{fig:adaptive_sample_size} shows that in DEEP and SimSearchNet++, the sample size was reduced while maintaining a low Q-error, demonstrating the efficiency of adaptive sampling. In contrast, SIFT required an increased sample size to achieve higher accuracy, highlighting the adaptive nature of \sysname in adjusting to dataset-specific characteristics. 




\begin{figure}[t]
    \centering
    \includegraphics[width=0.99\linewidth, trim={0pt 3pt 0pt 6pt},clip]{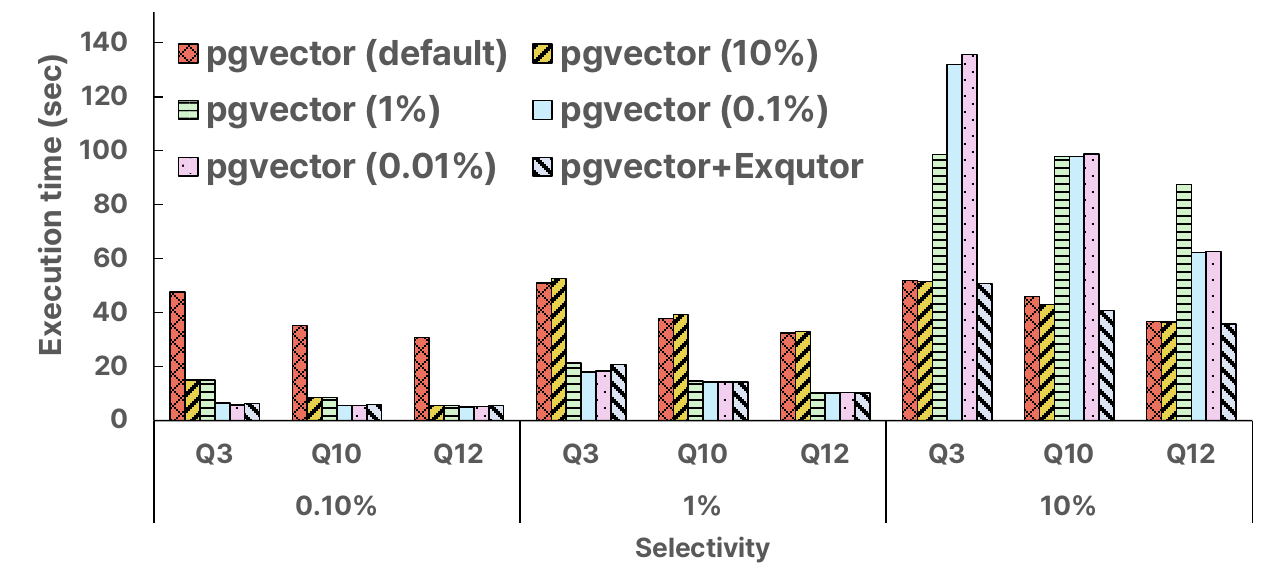}
    \caption{Query execution time for TPC-H VAQs without vector indexes under varying selectivities on the DEEP dataset with SF100 using pgvector and \sysname. The values in parentheses indicate the adjusted fixed selectivity parameter values used by pgvector’s heuristic selectivity.}
    \label{fig:query_time_selectivity}
    \vspace{-1mm}
\end{figure}

\mypar{Query time of sampling method across selectivity levels}
\change{
\autoref{fig:query_time_selectivity} shows the query execution time of both pgvector and \sysname under the sampling-based estimation method, evaluated at three different selectivity levels: 0.1\%, 1\%, and 10\%. As pgvector uses a static heuristic for cardinality estimation, its execution plans remain largely unchanged across different selectivity levels. This leads to inefficient plans and consistently high query latency, especially when selectivity is low. In contrast, \sysname dynamically adjusts its sampling size and accurately estimates cardinality, enabling the optimizer to generate plans that better reflect the selectivity of vector predicates. This allows \sysname to significantly outperform pgvector at low selectivities, where the benefit of avoiding over-scanning is most pronounced. At 10\% selectivity, the performance gap narrows because pgvector’s heuristic estimate approaches the actual selectivity, reducing the relative gain from sampling. 
}

\change{We further extended this analysis by varying the parameter shown in \autoref{tab:vss_magic_number} and comparing the resulting execution times. As shown in \autoref{fig:query_time_selectivity}, relying on a fixed sampling ratio fails to consistently yield optimal performance across different selectivities. The key determinant is the alignment between estimated and actual cardinalities, showing that a fixed parameter alone cannot ensure robust performance across varying selectivities.}

\begin{table}[t!]
  \centering
  \caption{Overhead and reduced execution time of ECQO and fixed sampling methods on pgvector for Query Q3 with DEEP dataset. The relative overhead indicates the ratio of optimization overhead to execution time reduction.}
  \label{tab:overhead}
  \setlength{\tabcolsep}{3pt} 
  \scalebox{0.9}{
  \begin{tabular}{lcccc}
    \hline
    \makecell{\textbf{Method}} & 
    \makecell{\textbf{Dataset}} & 
    \makecell{\textbf{Overhead} \\ \textbf{(ms)}} & 
    \makecell{\textbf{Reduced} \\ \textbf{Time (s)}} &
    \makecell{\textbf{Relative} \\ \textbf{Overhead (\%)}} \\
    \hline
    \multirowcell{3}[0pt][c]{\textbf{ECQO}}
      & DEEP                   & 1.88   & 43.34 & 0.0043 \\
      & SIFT                   & 1.89   & 41.65 & 0.0045 \\
      & \makecell{SimSearchNet++} & 1.96   & 43.66 & 0.0045 \\
    \hline
    \multirow{3}{*}{\makecell{\textbf{Fixed} \\ \textbf{Sampling}}} 
      & DEEP                   & 28.17  & 38.87 & 0.0724 \\
      & SIFT                   & 33.23  & 47.12 & 0.0705 \\
      & \makecell{SimSearchNet++} & 72.83  & 26.67 & 0.2730 \\
    \hline
  \end{tabular}
  }
  \vspace{-6mm}
\end{table}

\mypar{Overhead of \sysname}
\autoref{tab:overhead} presents the overhead introduced by ECQO and sampling-based estimation in \sysname for Q3 on the DEEP, SIFT, and SimSearchNet++ datasets. ECQO incurs minimal cost as it leverages a single lightweight traversal of the vector index during query planning. As shown, ECQO’s overhead is consistently low across all datasets, ranging from 1.88 to 1.96 milliseconds, while delivering substantial reductions in execution time, with savings between 41.65 and 43.66 seconds. The relative overhead remains extremely small, between 0.0043\% and 0.0045\%, indicating that ECQO offers high returns for negligible planning effort.
Sampling-based estimation introduces significant execution time savings, ranging from 26.67 to 47.12 seconds across the evaluated datasets. In return for these improvements, it introduces moderate planning overhead, measured between 28.17 and 72.83 ms, which arises from evaluating the similarity predicate on a sampled subset during query planning. The relative overhead remains low, from 0.0705\% to 0.273\%, with the highest value observed in the SimSearchNet++ dataset. This pattern is attributed to the higher computational cost of similarity evaluation in high-dimensional vector. Nevertheless, the overall overhead remains minimal in comparison to the achieved performance gains, validating the practicality of the approach.

\begin{figure}[t!]
    \centering
    \includegraphics[width=0.99\linewidth, trim={0pt 5pt 0pt 13pt},clip]{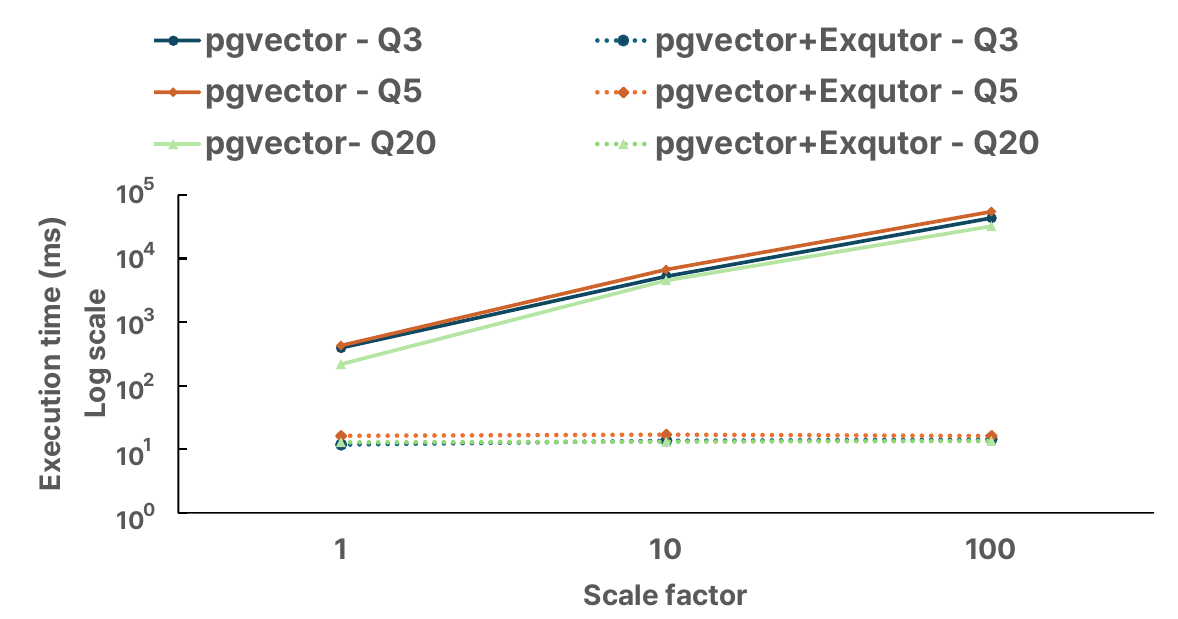}
    \caption{Scalability of \sysname(ECQO) on pgvector for TPC-H VAQs using the DEEP dataset, evaluated while increasing the scale factor.}
    \label{fig:pg_scalability}
    \vspace{-6mm}
\end{figure}

\shorten{
\mypar{Data scalability}
To evaluate the scalability of \sysname, we conducted experiments on the DEEP dataset using scale factors of 1, 10, and 100. \autoref{fig:pg_scalability} shows that pgvector experiences near-linear growth in execution time as dataset size increases.
In contrast, \sysname demonstrates highly stable performance across all scale factors by leveraging the property that HNSW search complexity does not grow linearly with data size.

Notably, all evaluated queries except Q20 exhibited similar trends, where \sysname consistently mitigated the impact of dataset scaling. In the case of Q20, execution time increases linearly due to the full scan of the \textit{lineitem} table. However, \sysname still achieves up to 1.31{\small$\times$} speedups, demonstrating that \sysname effectively mitigates the effect of increasing data size by leveraging accurate cardinality and efficient query planning, ensuring stable performance for large-scale VAQs.
}
\vspace{-1mm}
\change{
\subsection{Limitations}
\label{sec:limitations}
While \sysname demonstrates substantial performance improvements across diverse workloads, several limitations remain. 
In high-dimensional spaces, the overhead of sampling increases because of the higher cost of distance computations, which may reduce the efficiency of our adaptive sampling strategy. 
Moreover, our approach relies on cost models that fail to fully capture the performance characteristics of ANN indexes, so even with accurate cardinality estimates the optimizer may still choose suboptimal plans. 
Addressing these issues through more efficient sampling in high dimensions and refined cost models for VAQ optimization remains an important direction for future work.
}

\section{Related Work}

\mypar{Filtered vector search}
As vector similarity search becomes more prevalent, many systems~\cite{Milvus, Manu, qdrant} store vector embeddings alongside structured metadata, enabling filtered vector search. This trend has also emerged in ANN benchmarks~\cite{annbenchmark23, myscale-benchmark}, highlighting the growing importance of efficient filtering techniques. Several studies have optimized filtered vector queries by restructuring ANN indexes to support filtering constraints more effectively. ACORN~\cite{ACORN}, SeRF~\cite{serf}, HQANN~\cite{hqann}, and diskANN~\cite{diskANN} enhance ANN search by integrating attribute filtering directly into the index structure, improving retrieval efficiency. However, these methods are limited to filtering within a single relation or collection, making them less effective for large-scale analytical workloads that involve complex joins across multiple datasets.

\mypar{Query optimization in generalized vector database systems}
Several generalized vector database systems have extended traditional query processing techniques to support vector operations. AnalyticDB~\cite{AnalyticDB} optimizes filtered vector searches using a cost-based model, while SingleStore~\cite{singlestore} integrates filters directly into vector index scans to improve retrieval efficiency. However, these optimizations primarily target simple filter queries rather than complex analytical workloads involving multi-way joins and nested queries. As a result, they do not effectively address the challenges of optimizing VAQs, where inaccurate cardinality estimation can severely degrade query performance.

One technique in query optimization for efficiently estimating selectivity and cost is sampling. Early works introduced random sampling for join size estimation~\cite{random-sampling-joins, selectivity-cost-estimation}, while later approaches refined these ideas with adaptive sampling strategies~\cite{adaptive-sampling}. The method in~\cite{adaptive-sampling} adjusts the sample size dynamically until a desired confidence level is reached, but does not consider sampling overhead or optimize it dynamically based on query characteristics.

\section{conclusion}
In this paper, we introduce \sysname, an extended query optimizer designed to improve the performance of vector-augmented analytical queries by addressing the challenges of inaccurate cardinality estimation in vector searches. By leveraging exact cardinality query optimization and adaptive sampling, \sysname significantly enhances query performance, achieving speedups of up to four orders of magnitude. Through integration with pgvector, VBASE, and DuckDB, \sysname extends the ability of generalized vector database systems to efficiently handle vector-augmented analytical queries, contributing to the optimization of emerging data science pipelines like retrieval-augmented generation (RAG). This work demonstrates the critical role of accurate cardinality estimation and query optimization in enhancing the scalability and efficiency of modern data workflows.
\vspace{-1mm}
\section*{ACKNOWLEDGMENT}
This research was supported by NRF grants (No. RS-2025-16068623, No. RS-2024-NR121334), the Korea Basic Science Institute (National Research Facilities and Equipment Center) grant funded by the Ministry of Science and ICT (No. RS-2024-00403860), Advanced Database System Infrastructure (NFEC-2024-11-300458), and the Yonsei University Research Fund (2025-22-0057).

\bibliographystyle{IEEEtran}
\bibliography{reference}

@article{apple,
author = {Mohoney, Jason and Pacaci, Anil and Chowdhury, Shihabur Rahman and Mousavi, Ali and Ilyas, Ihab F. and Minhas, Umar Farooq and Pound, Jeffrey and Rekatsinas, Theodoros},
title = {High-Throughput Vector Similarity Search in Knowledge Graphs},
year = {2023},
issue_date = {June 2023},
publisher = {Association for Computing Machinery},
address = {New York, NY, USA},
volume = {1},
number = {2},
url = {https://doi.org/10.1145/3589777},
doi = {10.1145/3589777},
journal = {Proc. ACM Manag. Data},
month = jun,
articleno = {197},
numpages = {25}
}

@article{context-enhanced-realtional-operator, title={Context-Enhanced Relational Operators with Vector Embeddings}, url={http://arxiv.org/abs/2312.01476}, DOI={10.48550/arXiv.2312.01476},  note={arXiv:2312.01476}, number={arXiv:2312.01476}, publisher={arXiv}, author={Sanca, Viktor and Chatzakis, Manos and Ailamaki, Anastasia}, year={2023}, month=dec }

@misc{milvus_timeline,
  author = "{Milvus Team}",
  title = "{DNA Sequence Classification with Milvus}",
  howpublished = "\url{https://milvus.io/blog/2021-09-06-dna-sequence-classification-based-on-milvus.md}",
  note = "Accessed: 2024-11-30",
  year = {2024},
}

@misc{simsearchnet,
  author = "{Meta}",
  title = "{Here's how we're using AI to help detect misinformation}",
  howpublished = "\url{https://ai.meta.com/blog/heres-how-were-using-ai-to-help-detect-misinformation/}",
  note = "Accessed: 2024-11-30",
  year = {2020},
}

@inproceedings{diskANN,
author = {Gollapudi, Siddharth and Karia, Neel and Sivashankar, Varun and Krishnaswamy, Ravishankar and Begwani, Nikit and Raz, Swapnil and Lin, Yiyong and Zhang, Yin and Mahapatro, Neelam and Srinivasan, Premkumar and Singh, Amit and Simhadri, Harsha Vardhan},
title = {Filtered-DiskANN: Graph Algorithms for Approximate Nearest Neighbor Search with Filters},
year = {2023},
isbn = {9781450394161},
publisher = {Association for Computing Machinery},
url = {https://doi.org/10.1145/3543507.3583552},
doi = {10.1145/3543507.3583552},
booktitle = {Proceedings of the ACM Web Conference 2023},
pages = {3406–3416},
numpages = {11},
location = {Austin, TX, USA},
series = {WWW '23}
}

@inproceedings{ml_join,
  title={Analytical engines with context-rich processing: Towards efficient next-generation analytics},
  author={Sanca, Viktor and Ailamaki, Anastasia},
  booktitle={2023 IEEE 39th International Conference on Data Engineering (ICDE)},
  pages={3699--3707},
  year={2023},
  organization={IEEE}
}

@inproceedings{suv, address={New York, NY, USA}, series={SIGMOD ’18}, title={Accelerating Machine Learning Inference with Probabilistic Predicates}, ISBN={978-1-4503-4703-7}, url={https://doi.org/10.1145/3183713.3183751}, DOI={10.1145/3183713.3183751}, booktitle={Proceedings of the 2018 International Conference on Management of Data}, publisher={Association for Computing Machinery}, author={Lu, Yao and Chowdhery, Aakanksha and Kandula, Srikanth and Chaudhuri, Surajit}, year={2018}, month=may, pages={1493–1508}, collection={SIGMOD ’18} }

@article{serf,
author = {Zuo, Chaoji and Qiao, Miao and Zhou, Wenchao and Li, Feifei and Deng, Dong},
title = {SeRF: Segment Graph for Range-Filtering Approximate Nearest Neighbor Search},
year = {2024},
issue_date = {February 2024},
publisher = {Association for Computing Machinery},
volume = {2},
number = {1},
url = {https://doi.org/10.1145/3639324},
doi = {10.1145/3639324},
journal = {Proc. ACM Manag. Data},
month = mar,
articleno = {69},
numpages = {26}
}

@inproceedings{Vector-Database-Management,
author = {Pan, James Jie and Wang, Jianguo and Li, Guoliang},
title = {Vector Database Management Techniques and Systems},
year = {2024},
isbn = {9798400704222},
publisher = {Association for Computing Machinery},
url = {https://doi.org/10.1145/3626246.3654691},
doi = {10.1145/3626246.3654691},
booktitle = {Companion of the 2024 International Conference on Management of Data},
pages = {597–604},
numpages = {8},
series = {SIGMOD/PODS '24}
}

@INPROCEEDINGS{specialized-dbms,
  author={Zhang, Yunan and Liu, Shige and Wang, Jianguo},
  booktitle={2024 IEEE 40th International Conference on Data Engineering (ICDE)}, 
  title={Are There Fundamental Limitations in Supporting Vector Data Management in Relational Databases? A Case Study of PostgreSQL}, 
  year={2024},
  volume={},
  number={},
  pages={3640-3653},
  doi={10.1109/ICDE60146.2024.00280}}

@article{clickhouse,
author = {Schulze, Robert and Schreiber, Tom and Yatsishin, Ilya and Dahimene, Ryadh and Milovidov, Alexey},
title = {ClickHouse - Lightning Fast Analytics for Everyone},
year = {2024},
issue_date = {August 2024},
publisher = {VLDB Endowment},
volume = {17},
number = {12},
issn = {2150-8097},
url = {https://doi.org/10.14778/3685800.3685802},
doi = {10.14778/3685800.3685802},
journal = {Proc. VLDB Endow.},
month = nov,
pages = {3731–3744},
numpages = {14}
}

@inproceedings{vbase, title={{VBASE}: Unifying Online Vector Similarity Search and Relational Queries via Relaxed Monotonicity}, ISBN={978-1-939133-34-2}, url={https://www.usenix.org/conference/osdi23/presentation/zhang-qianxi}, author={Zhang, Qianxi and Xu, Shuotao and Chen, Qi and Sui, Guoxin and Xie, Jiadong and Cai, Zhizhen and Chen, Yaoqi and He, Yinxuan and Yang, Yuqing and Yang, Fan and Yang, Mao and Zhou, Lidong}, year={2023}, pages={377–395}, language={en} }

@ARTICLE{pq,
  author={Jégou, Herve and Douze, Matthijs and Schmid, Cordelia},
  journal={IEEE Transactions on Pattern Analysis and Machine Intelligence}, 
  title={Product Quantization for Nearest Neighbor Search}, 
  year={2011},
  volume={33},
  number={1},
  pages={117-128},
  doi={10.1109/TPAMI.2010.57}}

@ARTICLE{tree-aqp,
  author={Muja, Marius and Lowe, David G.},
  journal={IEEE Transactions on Pattern Analysis and Machine Intelligence}, 
  title={Scalable Nearest Neighbor Algorithms for High Dimensional Data}, 
  year={2014},
  volume={36},
  number={11},
  pages={2227-2240},
  doi={10.1109/TPAMI.2014.2321376}}

@article{tree-aqp-2,
author = {Lu, Kejing and Wang, Hongya and Wang, Wei and Kudo, Mineichi},
title = {VHP: approximate nearest neighbor search via virtual hypersphere partitioning},
year = {2020},
issue_date = {May 2020},
publisher = {VLDB Endowment},
volume = {13},
number = {9},
issn = {2150-8097},
url = {https://doi.org/10.14778/3397230.3397240},
doi = {10.14778/3397230.3397240},
journal = {Proc. VLDB Endow.},
month = may,
pages = {1443–1455},
numpages = {13}
}

@misc{annoy,
  author       = {Erik Bernhardsson},
  title        = {Annoy: Approximate Nearest Neighbors in C++/Python},
  year         = {2013},
  howpublished = {\url{https://github.com/spotify/annoy}},
  note         = {Accessed: 2024-12-01}
}

@article{hash,
author = {Zheng, Bolong and Zhao, Xi and Weng, Lianggui and Hung, Nguyen Quoc Viet and Liu, Hang and Jensen, Christian S.},
title = {PM-LSH: A fast and accurate LSH framework for high-dimensional approximate NN search},
year = {2020},
issue_date = {January 2020},
publisher = {VLDB Endowment},
volume = {13},
number = {5},
issn = {2150-8097},
url = {https://doi.org/10.14778/3377369.3377374},
doi = {10.14778/3377369.3377374},
journal = {Proc. VLDB Endow.},
month = jan,
pages = {643–655},
numpages = {13}
}

@article{hash3,
author = {Park, Yongjoo and Cafarella, Michael and Mozafari, Barzan},
title = {Neighbor-sensitive hashing},
year = {2015},
issue_date = {November 2015},
publisher = {VLDB Endowment},
volume = {9},
number = {3},
issn = {2150-8097},
url = {https://doi.org/10.14778/2850583.2850589},
doi = {10.14778/2850583.2850589},
journal = {Proc. VLDB Endow.},
month = nov,
pages = {144–155},
numpages = {12}
}

@article{hnsw,
  title={Efficient and robust approximate nearest neighbor search using hierarchical navigable small world graphs},
  author={Malkov, Yu A and Yashunin, Dmitry A},
  journal={IEEE transactions on pattern analysis and machine intelligence},
  volume={42},
  number={4},
  pages={824--836},
  year={2018},
  publisher={IEEE}
}

@article{graph,
author = {Fu, Cong and Xiang, Chao and Wang, Changxu and Cai, Deng},
title = {Fast approximate nearest neighbor search with the navigating spreading-out graph},
year = {2019},
issue_date = {January 2019},
publisher = {VLDB Endowment},
volume = {12},
number = {5},
issn = {2150-8097},
url = {https://doi.org/10.14778/3303753.3303754},
doi = {10.14778/3303753.3303754},
journal = {Proc. VLDB Endow.},
month = jan,
pages = {461–474},
numpages = {14}
}

@INPROCEEDINGS{graph2,
  author={Zhao, Weijie and Tan, Shulong and Li, Ping},
  booktitle={2020 IEEE 36th International Conference on Data Engineering (ICDE)}, 
  title={SONG: Approximate Nearest Neighbor Search on GPU}, 
  year={2020},
  volume={},
  number={},
  pages={1033-1044},
  doi={10.1109/ICDE48307.2020.00094}}

@misc{faiss,
  author       = {Facebook AI Research},
  title        = {Faiss: A library for efficient similarity search and clustering of dense vectors},
  year         = {2017},
  howpublished = {\url{https://github.com/facebookresearch/faiss}},
  note         = {Accessed: 2024-12-01}
}

@article{quantization,
  title={Billion-scale similarity search with GPUs},
  author={Johnson, Jeff and Douze, Matthijs and J{\'e}gou, Herv{\'e}},
  journal={IEEE Transactions on Big Data},
  volume={7},
  number={3},
  pages={535--547},
  year={2019},
  publisher={IEEE}
}

@misc{pgvector,
  author       = {Andrew Kane},
  title        = {pgvector: Open-source vector similarity search for PostgreSQL},
  year         = {2021},
  howpublished = {\url{http://github.com/pgvector}},
  note         = {Accessed: 2024-11-29}
}

@misc{chroma,
  author       = {Chroma},
  title        = {Chroma: The Open-Source AI Application Database},
  year         = {2024},
  howpublished = {\url{https://www.trychroma.com/}},
  note         = {Accessed: 2024-12-01}
}

@misc{pinecone,
  author       = {Pinecone Systems, Inc.},
  title        = {Pinecone: The Vector Database to Build Knowledgeable AI},
  year         = {2024},
  howpublished = {\url{https://www.pinecone.io/}},
  note         = {Accessed: 2024-12-01}
}

@misc{qdrant,
  author       = {Qdrant},
  title        = {Qdrant: High-Performance Vector Search at Scale},
  year         = {2024},
  howpublished = {\url{https://qdrant.tech/}},
  note         = {Accessed: 2024-12-01}
}

@inproceedings{duckdb,
  title={Duckdb: an embeddable analytical database},
  author={Raasveldt, Mark and M{\"u}hleisen, Hannes},
  booktitle={Proceedings of the 2019 international conference on management of data},
  pages={1981--1984},
  year={2019}
}

@misc{duckdb-vss,
  author       = {DuckDB Development Team},
  title        = {DuckDB-VSS: Vector Similarity Search Extension for DuckDB},
  year         = {2024},
  howpublished = {\url{https://github.com/duckdb/duckdb_vss}},
  note         = {Accessed: 2024-12-01}
}

@misc{vespa,
  author       = {Vespa.ai},
  title        = {Vespa: AI + Data, Online at Any Scale},
  year         = {2024},
  howpublished = {\url{https://vespa.ai/}},
  note         = {Accessed: 2024-12-01}
}

@misc{neo4j,
  author       = {Neo4j},
  title        = {Neo4j Vector Index and Search},
  year         = {2024},
  howpublished = {\url{https://neo4j.com/labs/genai-ecosystem/vector-search/}},
  note         = {Accessed: 2024-12-01}
}

@misc{redis,
  author       = {Redis},
  title        = {Redis for vector database},
  year         = {2024},
  howpublished = {\url{https://redis.io/solutions/vector-database/}},
  note         = {Accessed: 2024-12-01}
}

@misc{myscale-benchmark,
  author       = {myscale},
  title        = {vector-db-benchmark},
  year         = {2024},
  howpublished = {\url{https://github.com/myscale/vector-db-benchmark}},
  note         = {Accessed: 2024-12-01}
}

@article{annbenchmark23,
  title={Results of the Big ANN: NeurIPS'23 competition},
  author={Simhadri, Harsha Vardhan and Aum{\"u}ller, Martin and Ingber, Amir and Douze, Matthijs and Williams, George and Manohar, Magdalen Dobson and Baranchuk, Dmitry and Liberty, Edo and Liu, Frank and Landrum, Ben and others},
  journal={arXiv preprint arXiv:2409.17424},
  year={2024}
}

@article{survey-vector, title={Survey of vector database management systems}, volume={33}, ISSN={0949-877X}, DOI={10.1007/s00778-024-00864-x}, number={5}, journal={The VLDB Journal}, author={Pan, James Jie and Wang, Jianguo and Li, Guoliang}, year={2024}, month=sep, pages={1591–1615}, language={en} }

@article{singlestore,
  author    = {Cheng Chen and Chenzhe Jin and Yunan Zhang and Sasha Podolsky and Chun Wu and Szu{-}Po Wang and Eric Hanson and Zhou Sun and Robert Walzer and Jianguo Wang},
  title     = {{SingleStore-V: An Integrated Vector Database System in SingleStore}},
  journal={Proceedings of the VLDB Endowment},
  volume    = {17},
  number    = {12},
  pages     = {3772-3785},
  year      = {2024},
  url       = {https://doi.org/10.14778/3685800.3685805},
  doi       = {10.14778/3685800.3685805}
}

@misc{Manu,
      title={Manu: A Cloud Native Vector Database Management System}, 
      author={Rentong Guo and Xiaofan Luan and Long Xiang and Xiao Yan and Xiaomeng Yi and Jigao Luo and Qianya Cheng and Weizhi Xu and Jiarui Luo and Frank Liu and Zhenshan Cao and Yanliang Qiao and Ting Wang and Bo Tang and Charles Xie},
      year={2022},
      eprint={2206.13843},
      archivePrefix={arXiv},
      primaryClass={cs.DB},
      url={https://arxiv.org/abs/2206.13843}, 
}

@inproceedings{Milvus,
author = {Wang, Jianguo and Yi, Xiaomeng and Guo, Rentong and Jin, Hai and Xu, Peng and Li, Shengjun and Wang, Xiangyu and Guo, Xiangzhou and Li, Chengming and Xu, Xiaohai and Yu, Kun and Yuan, Yuxing and Zou, Yinghao and Long, Jiquan and Cai, Yudong and Li, Zhenxiang and Zhang, Zhifeng and Mo, Yihua and Gu, Jun and Jiang, Ruiyi and Wei, Yi and Xie, Charles},
title = {Milvus: A Purpose-Built Vector Data Management System},
year = {2021},
isbn = {9781450383431},
publisher = {Association for Computing Machinery},
url = {https://doi.org/10.1145/3448016.3457550},
doi = {10.1145/3448016.3457550},
booktitle = {Proceedings of the 2021 International Conference on Management of Data},
pages = {2614–2627},
numpages = {14},
series = {SIGMOD '21}
}

@article{AnalyticDB, title={AnalyticDB-V: a hybrid analytical engine towards query fusion for structured and unstructured data}, volume={13}, DOI={10.14778/3415478.3415541}, number={12}, journal={Proceedings of the VLDB Endowment}, author={Wei, Chuangxian and Wu, Bin and Wang, Sheng and Lou, Renjie and Zhan, Chaoqun and Li, Feifei and Cai, Yuanzhe}, year={2020}, month=aug, pages={3152–3165} }

@inproceedings{PASE, title={PASE: PostgreSQL Ultra-High-Dimensional Approximate Nearest Neighbor Search Extension}, url={https://dl.acm.org/doi/10.1145/3318464.3386131}, DOI={10.1145/3318464.3386131}, booktitle={Proceedings of the 2020 ACM SIGMOD International Conference on Management of Data}, publisher={ACM}, author={Yang, Wen and Li, Tao and Fang, Gai and Wei, Hong}, year={2020}, month=jun, pages={2241–2253} }

@inproceedings{momentum,
  title={On the importance of initialization and momentum in deep learning},
  author={Sutskever, Ilya and Martens, James and Dahl, George and Hinton, Geoffrey},
  booktitle={International conference on machine learning},
  pages={1139--1147},
  year={2013},
  organization={PMLR}
}

@article{selectivity,
  title={Selectivity estimation for range predicates using lightweight models},
  author={Dutt, Anshuman and Wang, Chi and Nazi, Azade and Kandula, Srikanth and Narasayya, Vivek and Chaudhuri, Surajit},
  journal={Proceedings of the VLDB Endowment},
  volume={12},
  number={9},
  pages={1044--1057},
  year={2019},
  publisher={VLDB Endowment}
}

@article{deepdb,
  title={Deepdb: Learn from data, not from queries!},
  author={Hilprecht, Benjamin and Schmidt, Andreas and Kulessa, Moritz and Molina, Alejandro and Kersting, Kristian and Binnig, Carsten},
  journal={arXiv preprint arXiv:1909.00607},
  year={2019}
}

@article{learned,
  title={Learned cardinalities: Estimating correlated joins with deep learning},
  author={Kipf, Andreas and Kipf, Thomas and Radke, Bernhard and Leis, Viktor and Boncz, Peter and Kemper, Alfons},
  journal={arXiv preprint arXiv:1809.00677},
  year={2018}
}

@inproceedings{deep,
  title={Efficient indexing of billion-scale datasets of deep descriptors},
  author={Babenko, Artem and Lempitsky, Victor},
  booktitle={Proceedings of the IEEE Conference on Computer Vision and Pattern Recognition},
  pages={2055--2063},
  year={2016}
}

@inproceedings{sift,
  title={Searching in one billion vectors: re-rank with source coding},
  author={J{\'e}gou, Herv{\'e} and Tavenard, Romain and Douze, Matthijs and Amsaleg, Laurent},
  booktitle={2011 IEEE International Conference on Acoustics, Speech and Signal Processing (ICASSP)},
  pages={861--864},
  year={2011},
  organization={IEEE}
}

@article{sample_size,
  title={Determining sample size},
  author={Israel, Glenn D and others},
  year={1992},
  publisher={University of Florida Cooperative Extension Service, Institute of Food and~…}
}

@inproceedings{hqann,
  title={HQANN: Efficient and robust similarity search for hybrid queries with structured and unstructured constraints},
  author={Wu, Wei and He, Junlin and Qiao, Yu and Fu, Guoheng and Liu, Li and Yu, Jin},
  booktitle={Proceedings of the 31st ACM International Conference on Information \& Knowledge Management},
  pages={4580--4584},
  year={2022}
}

@article{ACORN,
  title={Acorn: Performant and predicate-agnostic search over vector embeddings and structured data},
  author={Patel, Liana and Kraft, Peter and Guestrin, Carlos and Zaharia, Matei},
  journal={Proceedings of the ACM on Management of Data},
  volume={2},
  number={3},
  pages={1--27},
  year={2024},
  publisher={ACM New York, NY, USA}
}

@inproceedings{nhq, title={An Efficient and Robust Framework for Approximate Nearest Neighbor Search with Attribute Constraint}, volume={36}, url={https://proceedings.neurips.cc/paper_files/paper/2023/file/32e41d6b0a51a63a9a90697da19d235d-Paper-Conference.pdf}, booktitle={Advances in Neural Information Processing Systems}, publisher={Curran Associates, Inc.}, author={Wang, Mengzhao and Lv, Lingwei and Xu, Xiaoliang and Wang, Yuxiang and Yue, Qiang and Ni, Jiongkang}, editor={Oh, A. and Naumann, T. and Globerson, A. and Saenko, K. and Hardt, M. and Levine, S.}, year={2023}, pages={15738–15751} }

@inproceedings{random-sampling-joins,
  title={Random sampling over joins revisited},
  author={Zhao, Zhuoyue and Christensen, Robert and Li, Feifei and Hu, Xiao and Yi, Ke},
  booktitle={Proceedings of the 2018 International Conference on Management of Data},
  pages={1525--1539},
  year={2018}
}

@article{selectivity-cost-estimation,
  title={Selectivity and cost estimation for joins based on random sampling},
  author={Haas, Peter J and Naughton, Jeffrey F and Seshadri, S and Swami, Arun N},
  journal={Journal of Computer and System Sciences},
  volume={52},
  number={3},
  pages={550--569},
  year={1996},
  publisher={Elsevier}
}

@inproceedings{adaptive-sampling,
  title={Practical selectivity estimation through adaptive sampling},
  author={Lipton, Richard J and Naughton, Jeffrey F and Schneider, Donovan A},
  booktitle={Proceedings of the 1990 ACM SIGMOD international conference on Management of data},
  pages={1--11},
  year={1990}
}

@article{ecqo_relevent, title={Analyzing Query Optimizer Performance in the Presence and Absence of Cardinality Estimates}, url={http://arxiv.org/abs/2311.17293}, DOI={10.48550/arXiv.2311.17293}, note={arXiv:2311.17293 [cs]}, number={arXiv:2311.17293}, publisher={arXiv}, author={Datta, Asoke and Tsan, Brian and Izenov, Yesdaulet and Rusu, Florin}, year={2023}, month=nov }

@article{kepler, title={Kepler: Robust Learning for Parametric Query Optimization}, volume={1}, ISSN={2836-6573}, DOI={10.1145/3588963}, abstractNote={Most existing parametric query optimization (PQO) techniques rely on traditional query optimizer cost models, which are often inaccurate and result in suboptimal query performance. We propose Kepler, an end-to-end learning-based approach to PQO that demonstrates significant speedups in query latency over a traditional query optimizer. Central to our method is Row Count Evolution (RCE), a novel plan generation algorithm based on perturbations in the sub-plan cardinality space. While previous approaches require accurate cost models, we bypass this requirement by evaluating candidate plans via actual execution data and training anML model to predict the fastest plan given parameter binding values. Our models leverage recent advances in neural network uncertainty in order to robustly predict faster plans while avoiding regressions in query performance. Experimentally, we show that Kepler achieves significant improvements in query runtime on multiple datasets on PostgreSQL.}, number={1}, journal={Proceedings of the ACM on Management of Data}, author={Doshi, Lyric and Zhuang, Vincent and Jain, Gaurav and Marcus, Ryan and Huang, Haoyu and Altinbüken, Deniz and Brevdo, Eugene and Fraser, Campbell}, year={2023}, month=may, pages={1–25}, language={en} }

@inproceedings{ecqo_imma, address={Amsterdam Netherlands}, title={Exact Cardinality Query Optimization with Bounded Execution Cost}, ISBN={978-1-4503-5643-5}, url={https://dl.acm.org/doi/10.1145/3299869.3300087}, DOI={10.1145/3299869.3300087}, booktitle={Proceedings of the 2019 International Conference on Management of Data}, publisher={ACM}, author={Trummer, Immanuel}, year={2019}, month=jun, pages={2–17}, language={en} }

@inproceedings{ecqo_vec, address={Virtual Event China}, title={Learned Cardinality Estimation for Similarity Queries}, ISBN={978-1-4503-8343-1}, url={https://dl.acm.org/doi/10.1145/3448016.3452790}, DOI={10.1145/3448016.3452790}, booktitle={Proceedings of the 2021 International Conference on Management of Data}, publisher={ACM}, author={Sun, Ji and Li, Guoliang and Tang, Nan}, year={2021}, month=jun, pages={1745–1757}, language={en} }

@inproceedings{cardinality_important, address={New York, NY, USA}, series={SIGMOD ’04}, title={Robust query processing through progressive optimization}, ISBN={978-1-58113-859-7}, url={https://dl.acm.org/doi/10.1145/1007568.1007642}, DOI={10.1145/1007568.1007642}, booktitle={Proceedings of the 2004 ACM SIGMOD international conference on Management of data}, publisher={Association for Computing Machinery}, author={Markl, Volker and Raman, Vijayshankar and Simmen, David and Lohman, Guy and Pirahesh, Hamid and Cilimdzic, Miso}, year={2004}, month=jun, pages={659–670}, collection={SIGMOD ’04} }

@article{ce_important_microsoft, title={Analyzing the Impact of Cardinality Estimation on Execution Plans in Microsoft SQL Server}, volume={16}, ISSN={2150-8097}, DOI={10.14778/3611479.3611494}, author={Lee, Kukjin and Dutt, Anshuman and Narasayya, Vivek and Chaudhuri, Surajit}, year={2023}, month=jul, pages={2871–2883}, language={en} }

@inproceedings{Lucene,
  title={Vector search with OpenAI embeddings: Lucene is all you need},
  author={Xian, Jasper and Teofili, Tommaso and Pradeep, Ronak and Lin, Jimmy},
  booktitle={Proceedings of the 17th ACM International Conference on Web Search and Data Mining},
  pages={1090--1093},
  year={2024}
}

@article{tpc-h,
  title={New TPC benchmarks for decision support and web commerce},
  author={Poess, Meikel and Floyd, Chris},
  journal={ACM Sigmod Record},
  volume={29},
  number={4},
  pages={64--71},
  year={2000},
  publisher={ACM New York, NY, USA}
}

@inproceedings{blended_rag,
  title={Blended rag: Improving rag (retriever-augmented generation) accuracy with semantic search and hybrid query-based retrievers},
  author={Sawarkar, Kunal and Mangal, Abhilasha and Solanki, Shivam Raj},
  booktitle={2024 IEEE 7th International Conference on Multimedia Information Processing and Retrieval (MIPR)},
  pages={155--161},
  year={2024},
  organization={IEEE}
}

@inproceedings{scalable_springer, address={Cham}, title={Scalable Similarity Search for Big Data}, ISBN={978-3-319-16868-5}, abstractNote={Analysis of contemporary Big Data collections require an effective and efficient content-based access to data which is usually unstructured. This first implies a necessity to uncover descriptive knowledge of complex and heterogeneous objects to make them findable. Second, multimodal search structures are needed to efficiently execute complex similarity queries possibly in outsourced environments while preserving privacy. Four specific research objectives to tackle the challenges are outlined and discussed. It is believed that a relevant solution of these problems is necessary for a scalable similarity search operating on Big Data.}, booktitle={Scalable Information Systems}, publisher={Springer International Publishing}, author={Zezula, Pavel}, editor={Jung, Jason J. and Badica, Costin and Kiss, Attila}, year={2015}, pages={3–12} }

@ARTICLE{healthcare_rag_2,
  author={Su, Cheng and Wen, Jinbo and Kang, Jiawen and Wang, Yonghua and Su, Yuanjia and Pan, Hudan and Zhong, Zishao and Hossain, M. Shamim},
  journal={IEEE Internet of Things Journal}, 
  title={Hybrid RAG-Empowered Multi-Modal LLM for Secure Data Management in Internet of Medical Things: A Diffusion-Based Contract Approach}, 
  year={2024},
  volume={},
  number={},
  pages={1-1},
  doi={10.1109/JIOT.2024.3521425}}

@article{recommendation_rag_1,
  title={Tree-based RAG-Agent Recommendation System: A Case Study in Medical Test Data},
  author={Yang, Yahe and Huang, Chengyue},
  journal={arXiv preprint arXiv:2501.02727},
  year={2025}
}

@article{spider, title={Spider 2.0: Evaluating Language Models on Real-World Enterprise Text-to-SQL Workflows}, url={http://arxiv.org/abs/2411.07763}, DOI={10.48550/arXiv.2411.07763}, abstractNote={Real-world enterprise text-to-SQL workflows often involve complex cloud or local data across various database systems, multiple SQL queries in various dialects, and diverse operations from data transformation to analytics. We introduce Spider 2.0, an evaluation framework comprising 632 real-world text-to-SQL workflow problems derived from enterprise-level database use cases. The databases in Spider 2.0 are sourced from real data applications, often containing over 1,000 columns and stored in local or cloud database systems such as BigQuery and Snowflake. We show that solving problems in Spider 2.0 frequently requires understanding and searching through database metadata, dialect documentation, and even project-level codebases. This challenge calls for models to interact with complex SQL workflow environments, process extremely long contexts, perform intricate reasoning, and generate multiple SQL queries with diverse operations, often exceeding 100 lines, which goes far beyond traditional text-to-SQL challenges. Our evaluations indicate that based on o1-preview, our code agent framework successfully solves only 17.0% of the tasks, compared with 91.2% on Spider 1.0 and 73.0% on BIRD. Our results on Spider 2.0 show that while language models have demonstrated remarkable performance in code generation -- especially in prior text-to-SQL benchmarks -- they require significant improvement in order to achieve adequate performance for real-world enterprise usage. Progress on Spider 2.0 represents crucial steps towards developing intelligent, autonomous, code agents for real-world enterprise settings. Our code, baseline models, and data are available at https://spider2-sql.github.io.}, note={arXiv:2411.07763 [cs]}, number={arXiv:2411.07763}, publisher={arXiv}, author={Lei, Fangyu and Chen, Jixuan and Ye, Yuxiao and Cao, Ruisheng and Shin, Dongchan and Su, Hongjin and Suo, Zhaoqing and Gao, Hongcheng and Hu, Wenjing and Yin, Pengcheng and Zhong, Victor and Xiong, Caiming and Sun, Ruoxi and Liu, Qian and Wang, Sida and Yu, Tao}, year={2024}, month=nov }

@article{knn-1,
  title={K-nearest neighbor},
  author={Peterson, Leif E},
  journal={Scholarpedia},
  volume={4},
  number={2},
  pages={1883},
  year={2009}
}

@article{knn-2,
  title={A review of various k-nearest neighbor query processing techniques},
  author={Dhanabal, Subramaniam and Chandramathi, SJIJCA},
  journal={International Journal of Computer Applications},
  volume={31},
  number={7},
  pages={14--22},
  year={2011},
  publisher={Citeseer}
}

@article{knn-3,
  title={Generalized k-nearest neighbor rules},
  author={Bezdek, James C and Chuah, Siew K and Leep, David},
  journal={Fuzzy Sets and Systems},
  volume={18},
  number={3},
  pages={237--256},
  year={1986},
  publisher={Elsevier}
}

@article{errors-in-cardinality,
  title={A survey on advancing the dbms query optimizer: Cardinality estimation, cost model, and plan enumeration},
  author={Lan, Hai and Bao, Zhifeng and Peng, Yuwei},
  journal={Data Science and Engineering},
  volume={6},
  pages={86--101},
  year={2021},
  publisher={Springer}
}

@article{tpch-quantifying,
  title={Quantifying TPC-H choke points and their optimizations},
  author={Dreseler, Markus and Boissier, Martin and Rabl, Tilmann and Uflacker, Matthias},
  journal={Proceedings of the VLDB Endowment},
  volume={13},
  number={8},
  pages={1206--1220},
  year={2020},
  publisher={VLDB Endowment}
}

@misc{andb,
      title={AnDB: Breaking Boundaries with an AI-Native Database for Universal Semantic Analysis}, 
      author={Tianqing Wang and Xun Xue and Guoliang Li and Yong Wang},
      year={2025},
      eprint={2502.13805},
      archivePrefix={arXiv},
      primaryClass={cs.DB},
      url={https://arxiv.org/abs/2502.13805}, 
}

@article{ann_index_large_size,
  title={Characterizing the dilemma of performance and index size in billion-scale vector search and breaking it with second-tier memory},
  author={Cheng, Rongxin and Peng, Yifan and Wei, Xingda and Xie, Hongrui and Chen, Rong and Shen, Sijie and Chen, Haibo},
  journal={arXiv preprint arXiv:2405.03267},
  year={2024}
}

@article{ecqo_test,
  title={Exact cardinality query optimization for optimizer testing},
  author={Chaudhuri, Surajit and Narasayya, Vivek and Ramamurthy, Ravi},
  journal={Proceedings of the VLDB Endowment},
  volume={2},
  number={1},
  pages={994--1005},
  year={2009},
  publisher={VLDB Endowment}
}

@inproceedings{selnet,
  title={Consistent and flexible selectivity estimation for high-dimensional data},
  author={Wang, Yaoshu and Xiao, Chuan and Qin, Jianbin and Mao, Rui and Onizuka, Makoto and Wang, Wei and Zhang, Rui and Ishikawa, Yoshiharu},
  booktitle={Proceedings of the 2021 International Conference on Management of Data},
  pages={2319--2327},
  year={2021}
}

@article{yfcc,
  title={Yfcc100m: The new data in multimedia research},
  author={Thomee, Bart and Shamma, David A and Friedland, Gerald and Elizalde, Benjamin and Ni, Karl and Poland, Douglas and Borth, Damian and Li, Li-Jia},
  journal={Communications of the ACM},
  volume={59},
  number={2},
  pages={64--73},
  year={2016},
  publisher={ACM New York, NY, USA}
}

@article{data-distribution,
  title={PDX: A Data Layout for Vector Similarity Search},
  author={Kuffo, Leonardo and Krippner, Elena and Boncz, Peter},
  journal={Proceedings of the ACM on Management of Data},
  volume={3},
  number={3},
  pages={1--26},
  year={2025},
  publisher={ACM New York, NY, USA}
}

@inproceedings{tpc-ds_1,
  title={The making of TPC-DS.},
  author={Nambiar, Raghunath Othayoth and Poess, Meikel},
  booktitle={VLDB},
  volume={6},
  pages={1049--1058},
  year={2006}
}

@inproceedings{tpc-ds_2,
  title={Why You Should Run TPC-DS: A Workload Analysis.},
  author={Poess, Meikel and Nambiar, Raghunath Othayoth and Walrath, David},
  booktitle={VLDB},
  volume={7},
  pages={1138--1149},
  year={2007}
}

@article{overestimation_problem,
  title={Statadvisor: Recommending statistical views},
  author={El-Helw, Amr and Ilyas, Ihab F and Zuzarte, Calisto},
  journal={Proceedings of the VLDB Endowment},
  volume={2},
  number={2},
  pages={1306--1317},
  year={2009},
  publisher={VLDB Endowment}
}

@misc{huggingface-wikipedia,
  title        = {Wikipedia (December 2022) Dataset},
  author       = {Cohere},
  howpublished = {\url{https://huggingface.co/datasets/Cohere/wikipedia-22-12}},
  note         = {Accessed: 2025-09-20}
}

\renewcommand{\thefigure}{L\arabic{figure}}
\renewcommand{\theHfigure}{L.\arabic{figure}}
\renewcommand{\thetable}{L\arabic{table}}
\renewcommand{\theHtable}{L.\arabic{table}}
\setcounter{figure}{0}
\setcounter{table}{0}

\end{document}